\documentclass{emulateapj}
\usepackage{epsfig,psfig,lscape}

\newcommand{\mdot}{\dot{M}}
\newcommand{\lsun}{{L}_{\odot}}
\newcommand{\msun}{{M}_{\odot}}
\newcommand{\rsun}{{R}_{\odot}}
\newcommand{\msyr}{\msun \ {\rm yr^{-1}}}

\slugcomment{}

\shorttitle{A ``Combination Nova'' in Z Andromedae}
\shortauthors{Sokoloski et al.}

\begin{document}


\title{A ``Combination Nova'' Outburst in Z Andromedae: Nuclear Shell Burning
Triggered by a Disk Instability}


\author{J. L. Sokoloski\altaffilmark{1,2},
S. J. Kenyon\altaffilmark{1}, B. R. Espey\altaffilmark{3,4},
Charles D. Keyes\altaffilmark{5}, S. R. McCandliss\altaffilmark{6},
A. K. H. Kong\altaffilmark{1}, J. P. Aufdenberg\altaffilmark{7},
A. V. Filippenko\altaffilmark{8}, W. Li\altaffilmark{8},
C. Brocksopp\altaffilmark{9}, Christian R. Kaiser\altaffilmark{10},
P. A. Charles\altaffilmark{10,11}, M. P. Rupen\altaffilmark{12}, and
R. P. S. Stone\altaffilmark{13}} 


\altaffiltext{1}{Smithsonian Astrophysical Observatory, 60 Garden
Street, Cambridge, MA 02138} 
\altaffiltext{2}{NSF Astronomy \& Astrophysics Fellow}
\altaffiltext{3}{School of Physics, Trinity College Dublin, Dublin 2,
Ireland}
\altaffiltext{4}{School of Cosmic Physics, Dublin Institute for
Advanced Studies, 5 Merrion Square, Dublin 2, Ireland}  
\altaffiltext{5}{Space Telescope Science Institute, 3700 San Martin
Dr., Baltimore, MD 21218}
\altaffiltext{6}{The Johns Hopkins University, Department of Physics and
Astronomy, 3400 N. Charles St., Baltimore, MD 21218}
\altaffiltext{7}{NOAO, P. O. Box 26732, Tucson, AZ 85726}
\altaffiltext{8}{Astronomy Department, 601 Campbell Hall, University
of California, Berkeley, CA 94720}
\altaffiltext{9}{Mullard Space Science Laboratory, University College London,
Dorking, Surry RH5 6NT}
\altaffiltext{10}{University of Southampton, Southampton SO17 1BJ, U.K.}
\altaffiltext{11}{South African Astronomical Observatory, P. O. Box 9, 
Observatory, 7935, South Africa}
\altaffiltext{12}{NRAO, P.~O.~Box 0, 1003 Lopezvill Rd, Socorro, NM 87801}
\altaffiltext{13}{UCO/Lick Observatory, Mt. Hamilton, CA 95140}


\begin{abstract}

We describe observational evidence for a new kind of
interacting-binary-star outburst that involves both an accretion
instability and an increase in thermonuclear shell burning on the
surface of an accreting white dwarf.  We refer to this new type of
eruption as a {\it combination nova}.  In late 2000, the prototypical
symbiotic star Z Andromedae brightened by roughly two magnitudes in
the optical.  We observed the outburst in the radio with the VLA and
MERLIN, in the optical both photometrically and spectroscopically, in
the far ultraviolet with $FUSE$, and in the X-rays with both $Chandra$
and $XMM$.  The two-year-long event had three distinct stages.  During
the first stage, the optical rise closely resembled an earlier, small
outburst that was caused by an accretion-disk instability. In the
second stage, the hot component ejected an optically thick shell of
material.  In the third stage, the shell cleared to reveal a white
dwarf whose luminosity remained on the order of $10^4\,
\lsun$ for approximately one year.  The eruption was thus too energetic
to have been powered by accretion alone.  We propose that the initial
burst of accretion was large enough to trigger enhanced nuclear
burning on the surface of the white dwarf and the ejection of an
optically thick shell of material.  This outburst therefore combined
elements of both a dwarf nova and a classical nova.  Our results have
implications for the long-standing problem of producing shell flashes
with short recurrence times on low-mass white dwarfs in symbiotic
stars.

\end{abstract}




\section{Introduction}

\subsection{Symbiotic-Star Outbursts} \label{sec:intross}

Symbiotic stars are interacting binaries in which material is
transferred from an evolved red-giant star to a more compact, hot
star, usually a white dwarf \citep[WD; see, e.g.,][and references
therein]{kbook, lapalmabook}.  In most symbiotics, the red giant
under-fills its Roche lobe, and the mass transfer proceeds via
gravitational capture of the red giant's wind.  An accretion disk may
or may not form
\citep[depending primarily on the relative velocity of the red-giant
wind and the accreting white dwarf; 
][]{livio88}.  Radiation from the accreting WD partially
ionizes the nebula formed by the red-giant wind, and this ionized
nebula gives rise to optical through far ultraviolet (FUV) emission
lines \citep{kbook,mur91,espey96,espey03,nuss03}.

The hot components in symbiotic stars typically have luminosities of
around $10^3\, \lsun$ \citep{mur91}.  To produce such a high
luminosity by accretion onto the WD alone, the accretion rate would
need to be $\sim 10^{-6} \msyr$.  But it is difficult for the WD to
accrete from the red-giant wind at this rate since red giants in
symbiotics typically lose mass at $\sim 10^{-7}$ $\msyr$ \citep{st90}.
In addition, $10^{-6} \msyr$ is well above the accretion rate needed
to produce quasi-steady thermonuclear shell burning
\citep[][]{paczyt78, sionea79, pacrudak80,iben82,sionstar86,
sionready92, sionstar94}. 
\citet{sbh01} found that 
stochastic optical flickering 
from an accretion disk, like that seen in
cataclysmic variable stars, is 
undetectable in the vast majority symbiotics -- probably because
it is hidden by nuclear-shell-burning light that has been reprocessed
into the optical by the nebula.  In fact, the inferred luminosities of
$\sim 10^3\, \lsun$ are easily produced by the nuclear burning of
hydrogen-rich material at the rate of a few times $10^{-8}\
\msyr$.  Accretion onto the WD in a typical symbiotic can supply fuel 
at this rate \citep{iijima02}.
Thus, some level of quasi-steady nuclear shell burning on the surface
of the WD appears to be a common feature of symbiotic systems.

Symbiotic stars also experience several different types of outbursts.  
The most common type of outburst, termed a classical symbiotic
outburst, recurs roughly every decade or so and has optical
amplitudes of several mag \citep{kbook}.  The nature of these
eruptions is not yet known.
Their peak luminosities appear to be too high for them to be
disk instabilities like those of dwarf novae
\citep{kbook,mskaea95}.  
They also recur too frequently to be nova-like thermonuclear
runaways
those in RS Oph and T CrB
\citep{iben82,fujimoto82,kt83,starrfieldea85,livioea89,mskaea95} or
even weak shell flashes \citep{sionea79}.
The fundamental difficulty is that most symbiotics contain low-mass
WDs.  Short recurrence times can only be produced for thermonuclear
runaways on WDs whose masses are very close to the Chandrasekhar
limit.  For example, \cite{sionea79} found that even a 1.2-$\msun$ WD
accreting at $10^{-8}\, \msyr$ required more than 600 yr between
shell flashes.  Thus, some additional physics must be involved
in classical symbiotic outbursts.
Since the mass loss during classical symbiotic outbursts can take the
form of a collimated jet
\citep[e.g.,][]{kellogg01,crocker01,brock04}, the nature of these
outbursts is also linked to the issue of jet formation.

Classical symbiotic-star optical outbursts are
tied to changes in the hot component of the binary.
They can therefore be
the optical brightening of either the accretion disk, the white dwarf,
or both \citep{kw84}.  An accretion disk brightens when mass transfer
through the disk increases, either because more material is being fed
to the disk or
because of an 
accretion-disk instability.  The white dwarf can brighten in the
optical due to either a decrease in effective temperature at constant
bolometric luminosity, which
shifts the WD spectral energy distribution to lower energies, or an
increase in the nuclear burning rate on the white-dwarf surface. A
factor that complicates this picture slightly is the nebula, which
reprocesses high-energy photons into the optical, where most outbursts
are primarily observed.  Symbiotic stars can therefore also brighten
in the optical due to an increase in reprocessed high-energy photons
\citep[see][for a model of the reprocessing]{nussvogel89}. 

Existing models for classical symbiotic outbursts include expansion of
the WD photosphere at constant bolometric luminosity due to the
accretion rate rising above the maximum value for steady burning
\citep{tutuyun,iben82}, a thermal pulse or shell flash \citep{kt83}, or a
dwarf-nova-like accretion-disk instability
\citep{duschl86a,duschl86b,mska02}.  The photospheric-expansion model has
shown promise for supersoft X-ray sources,
where it explains the inverse relationship between optical and X-ray
fluxes 
seen in some objects
\citep[e.g., RX J0513-69;][]{southwell96},  
which arises from changes in the location of the peak of the black-body
spectral energy distribution.  Both the thermal-pulse and
accretion-disk-instability models are related to phenomena that occur
in cataclysmic variables (CVs) --- runaway nuclear burning in
classical or recurrent novae, and accretion-disk limit cycles in dwarf
novae.
 
\subsection{Z Andromedae} \label{sec:intro_zand}

Z Andromedae (Z And) is the prototype for the class of symbiotic stars
\citep{kbook,ssproc88,lapalmabook}.  It has an orbital period of 759 d
\citep{formleib94,mk96}, 
orbital inclination and orientation angles of $47 \pm 12^{\circ}$ and
$72 \pm 6^{\circ}$ respectively \citep{schsch97}, and a WD mass of
$0.65 \pm 0.28\, \msun$ \cite[][for a total system mass of between 1.3 and 2.3
$\msun$]{schsch97}.  Distance estimates 
for Z And range from 0.6~kpc to 2.19~kpc, with an average value of
$1.2 \pm 0.5$~kpc \citep[see][and references therein]{kennyphd}.

Zandstra-method estimates of the quiescent WD luminosity
give $L_{hot} \approx 10^3\,\lsun$ \citep{mur91},
where $L_{hot}$ is the WD (hot component) luminosity.  It is therefore
very likely that some degree of nuclear shell burning is taking place
on the WD surface.  Z And is also the only symbiotic star for which
coherent optical oscillations at the WD spin period of 28 minutes
convincingly indicate the presence of a strongly magnetic WD
\citep{sb99}.  It was the first binary known to contain a WD
with both a strong magnetic field and quasi-steady nuclear shell
burning.  Finally, Z And is one of a small ($\sim$ 10, but growing)
number of symbiotic systems 
known to produce collimated
outflows in the form of non-relativistic jets \citep{brock04}.

The long-term light curve of Z And shows many examples of classical
symbiotic outbursts \cite[see][p.~4]{kbook} and demonstrates the
non-uniformity of this poorly defined type of eruption.  The outbursts
sometimes come in pairs (as in 1984 and 1986), or in a series of
eruptions with decreasing maximum brightnesses 
separated in time by slightly less than the orbital period
\citep{kw84,formleib94}.  In addition to having different sizes, the
outbusts in Z And also have different shapes, with some having a
steeper rise than decline (as in 1997 and comparable to dwarf-nova
light curves), others having a rise that occurs at roughly the same
pace as the decay (such as in 1984), and still others having more
complex shapes.
In the 2000--2002 event, the size of the
outburst ($\Delta V 
\approx 2$ mag) was typical for the system, too large
to be a simple disk instability, and much smaller than outbursts of
the symbiotic recurrent novae.  It was instead closer to the $\sim$3
mag of a typical symbiotic (slow) nova, as in AG Peg
\citep{kpk01}.

\citet{fc95} demonstrated the value of
multi-wavelength observations of classical symbiotic outbursts with
their study of the 1984--1986 pair of outbursts of Z And.  They found
that a shell of material was ejected during each of these two
outbursts, and that the overall behavior of the UV
lines and radio fluxes was consistent with an
accretion-rate fluctuation pushing the system over the maximum
accretion rate for quasi-steady nuclear burning.
\citet[][]{sb99}, however, detected
a 28-minute oscillation from magnetic hot spots on the WD surface
throughout the small outburst in 1997 and concluded that the WD
surface was not hidden by a shell ejection during that event.  They
suggested that it could instead have been due to an accretion-disk
instability.

\begin{figure*}[t]
\hspace{-0.5cm}
\epsfig{file=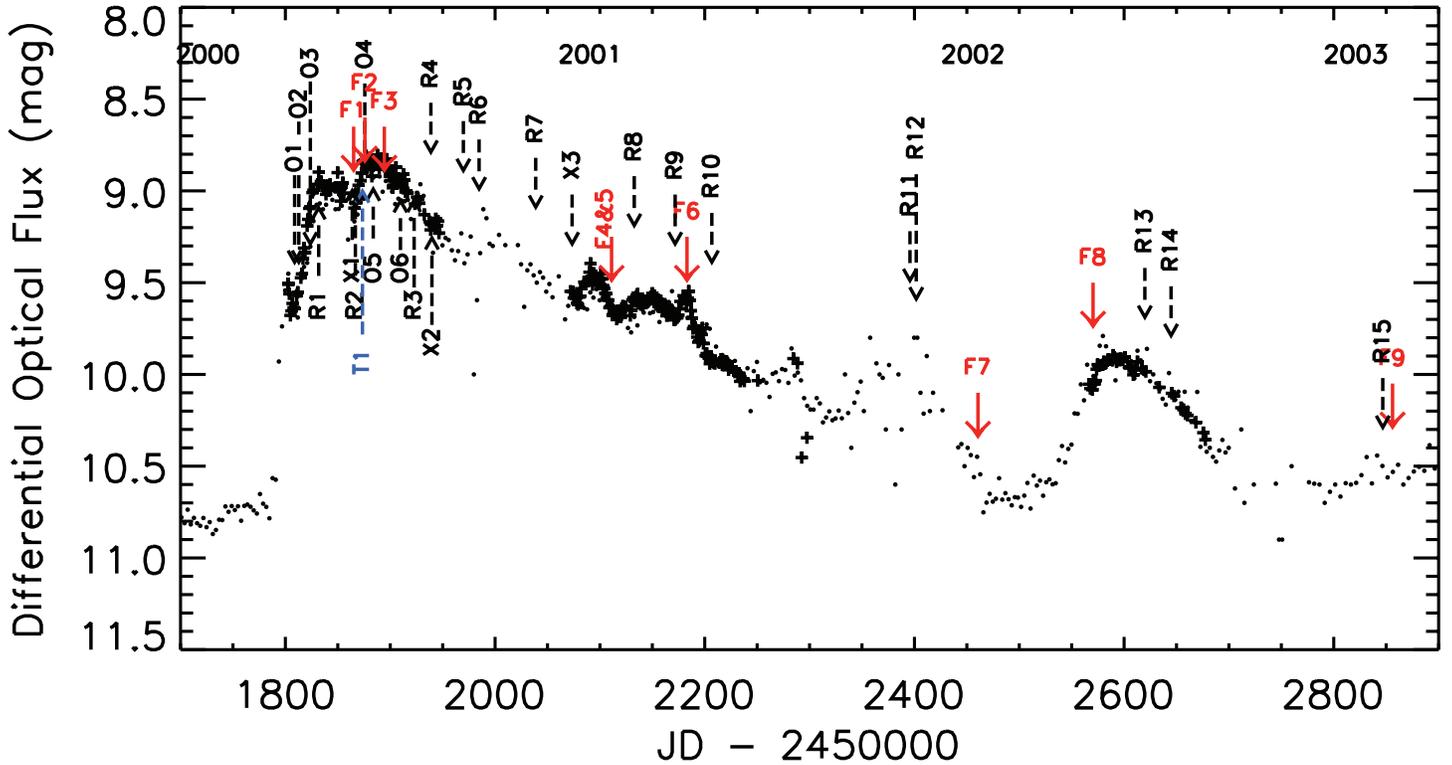,width=7.5in}
\caption{Long-term $V$-band light curve from the KAIT monitoring
described in \S\ref{sec:obs_ltlcs} (crosses) and the American
Association of Variable Star Observers (AAVSO; dots), with the times
of other observations marked.  O1--O6 refer to selected optical spectroscopic
observations listed in Table~\ref{tab:optspec}.
F1--F9 refer to FUV observations listed in Table~\ref{tab:fuv}.
R1--R15 refer to the radio observations listed in
Table~\ref{tab:radio}.  X1--X3 refer to X-ray observations listed in
Table~\ref{tab:xray}.  T1 
refers to the fast photometric timing
search described in
\S\ref{sec:obs_rapidvar}.
\label{fig1}}
\end{figure*}

To further investigate the nature and cause of classical
symbiotic-star outbursts, we 
obtained multi-wavelength
data, including observations with the $FUSE$, $XMM$, and $Chandra$
satellites, the VLA and MERLIN radio interferometers, and ground-based
optical spectroscopy and photometry, during the recent 2000--2003
activity phase of Z And.  We describe these observations and results
in \S\ref{sec:obs}.  In \S\ref{sec:analysis}, we use the
data to estimate the effective-temperature and luminosity evolution of
the WD,
and to investigate the ejection of a shell of material associated with
the activity in 2000--2001.  The WD temperatures and luminosities help
us determine that the 1997 outburst was due to an
accretion-disk instability similar to those of dwarf novae, and that
the 2000--2002 event was due to both an accretion-disk instability and
an increase in the rate of shell burning on the WD.  In
\S\ref{sec:implications}, we propose a {\it combination nova} model
for the 2000--2002 outburst of Z And 
where the 2000-2003 activity period was triggered by an
accretion-disk instability, but then primarily powered by the
resulting increase in nuclear burning on the surface of the white
dwarf.
We
summarize our findings in
\S\ref{sec:sum}.  UT dates are used throughout this paper.

\section{Observations and Results} 
\label{sec:obs} 

Although the hot components (e.g., white-dwarf plus accretion disk) in
symbiotic stars emit most of their energy in the FUV,
they also radiate significantly at radio through X-ray wavelengths.
Important diagnostics are found in each of these observational
regimes.  To make progress on the long-standing problem of the
fundamental nature and cause of classical symbiotic-star outbursts, we
observed the 2000--2002 outburst of Z And in the radio, optical, FUV,
and X-rays.
A summary of the timing and coverage of our data is
shown in Fig.~\ref{fig1}.  We provide a
listing of the 
observations 
in Tables~\ref{tab:optphot}--\ref{tab:xray}.

\subsection{$UBV$ Nightly Monitoring} \label{sec:obs_ltlcs}

\begin{figure*}[t]
\epsfig{file=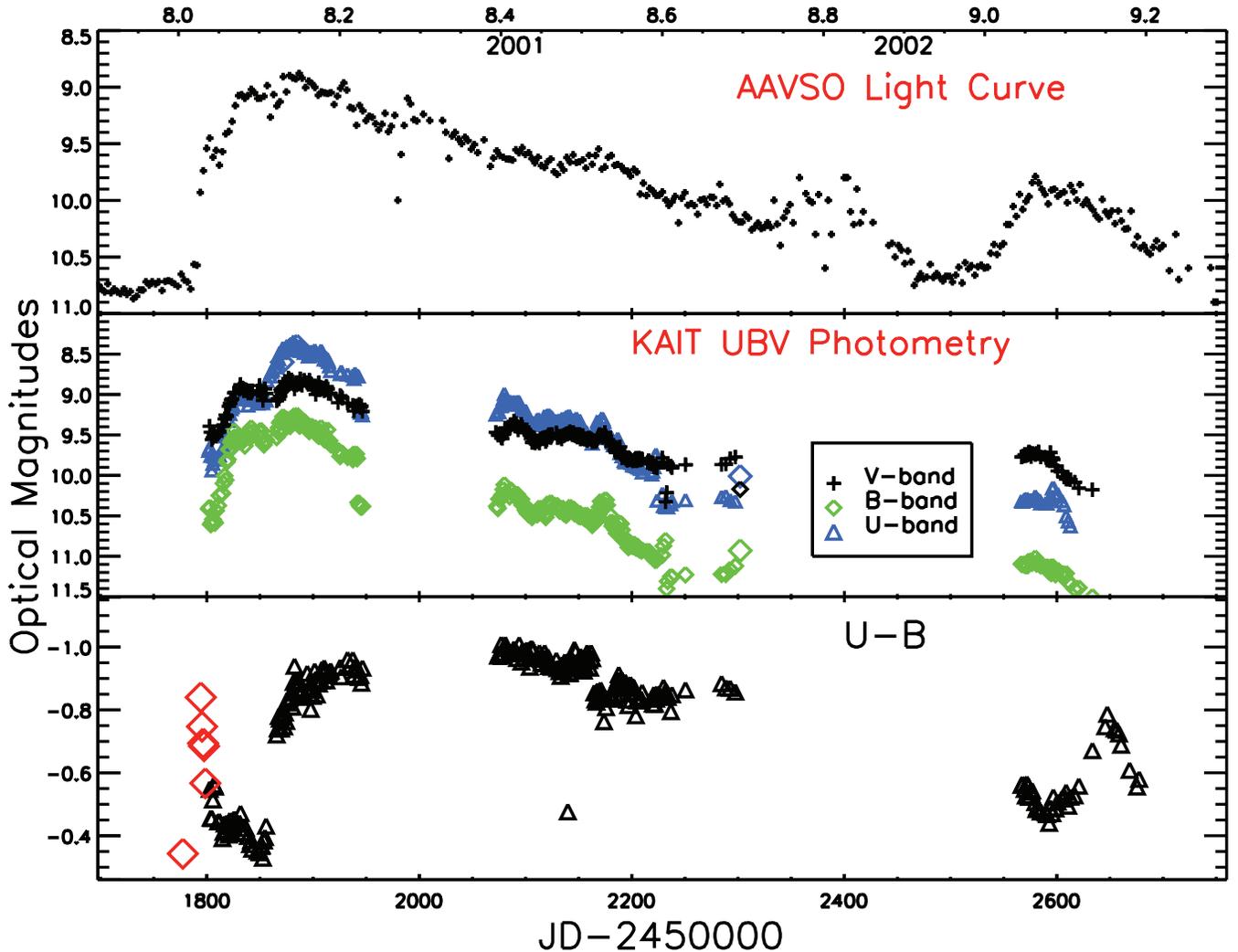,width=7in}
\caption{Light curves of Z And during the 2000--2002 outburst. Top
panel: full visual light curve from the AAVSO. Middle panel:
differential KAIT light curves, normalized to the absolute values of
\cite{tomov03}.  The error bars are smaller than the plot symbols.
Bottom panel: $U-B$ color.  Data taken by \cite{sko02} before we began
our KAIT monitoring are included (large open diamonds) to show the
blue color spike during the first stage of the optical brightening.
The first $U-B$ point, from
\cite{sko02}, is from approximately two weeks before the beginning of the
outburst. Orbital phase from \cite{mk96} is plotted on the top abscissa.
\label{fig:kaitfig}}
\end{figure*}

$UBV$ observations were taken nightly starting on JD 2451802 (2000
September 14)
with the 0.76-m Katzman Automatic Imaging Telescope
\citep[KAIT;][]{wli2000,filip2001,filip2005} at 
UCO/Lick Observatory on Mt. Hamilton near San Jose, CA.  Z And is
observable with KAIT from about mid-year through January, so the KAIT
light curves are divided into three portions from the latter parts of
2000, 2001, and 2002 (see Fig.~\ref{fig:kaitfig}). With integration
times ranging from 45 to 60 s for the $U$ band, 15 to 20 s for the $B$
band, and 5 to 10 s for the $V$ band, we typically obtained
signal-to-noise ratios (S/N) of greater than 100 for differential photometry
with respect to two comparison stars in the field of Z And.
The KAIT light curves plotted in Fig.~\ref{fig:kaitfig} have been
converted from the KAIT $UBV$ filter system to the standard Johnson
system 
and normalized using the absolute photometry of \cite{tomov03} and
\cite{sko02}.  We list the results of the photometric monitoring in
Table~\ref{tab:optphot}. 

\begin{figure*}[t]
\epsfig{file=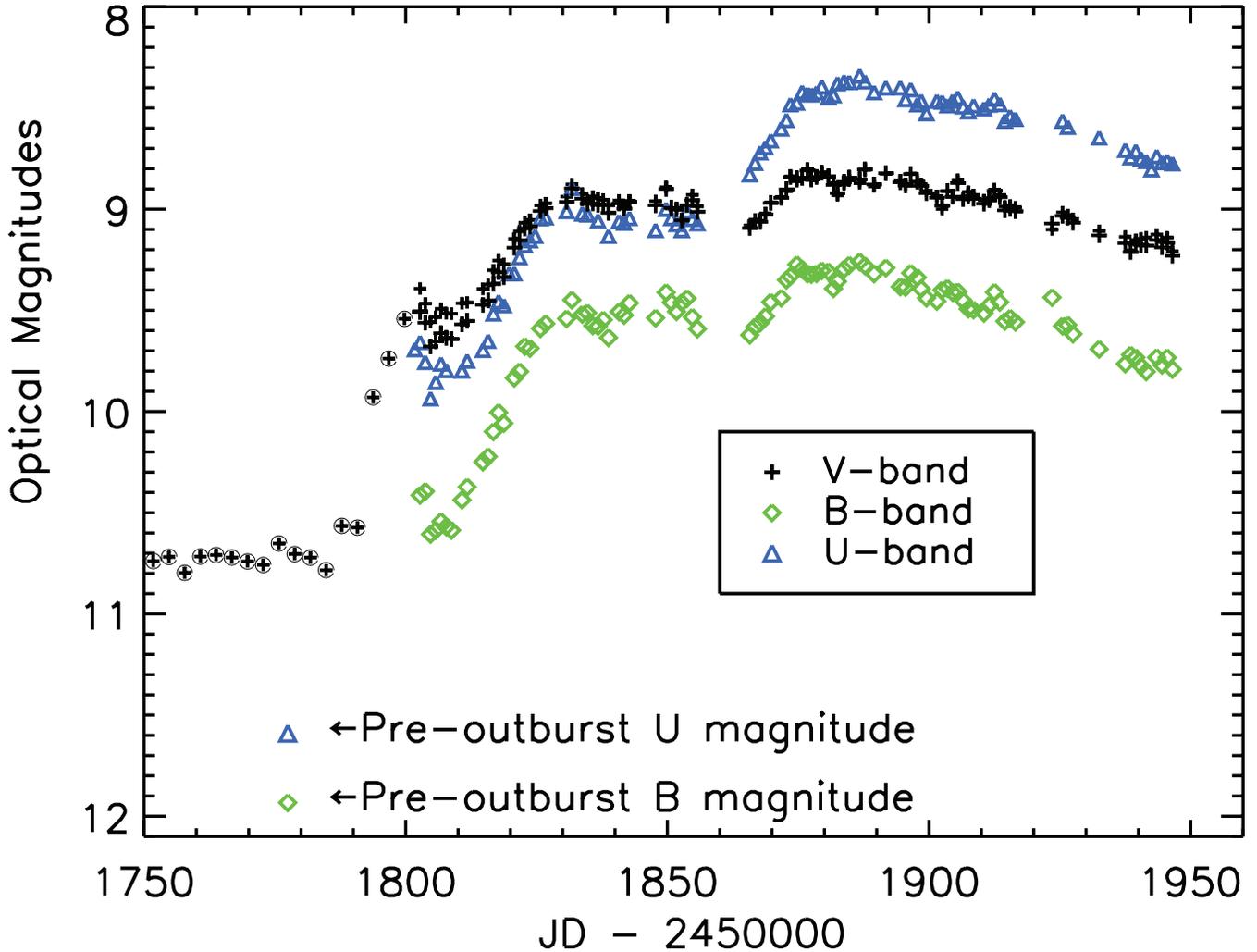,width=7in}
\caption{KAIT $V$-band (crosses), $B$-band (diamonds), and $U$-band
(triangles) nightly photometry reveals the stair-step or three-stage
nature of the rise to optical maximum. $V$-band data
from before the outburst and
during the initial fast rise (circled crosses) are from the AAVSO.
The pre-outburst $U$ and $B$ points are from \cite{sko02}.
\label{fig:kcloseup}} 
\end{figure*}

The high quality of the KAIT light curves immediately reveals 
that 
the rise to optical maximum in 2000
had three distinct
bursts, separated by two plateau periods.  The first, second, and
third optical rises took approximately 
2.5, 2.5--3, and 
slightly more than 3 weeks, respectively, and the
two plateaus lasted for about 1 week and then about 1 month respectively.
Fig.~\ref{fig:kcloseup} is a close-up view of the first 150 days of
the 2000--2002 outburst, which clearly shows the three stages.

Combined with 
photometry from the American Association of Variable Star Observers
(AAVSO) and \cite{sko02} taken just before the beginning of the
2000--2002 outburst,
the KAIT data show that the initial color evolution of the system also
had three distinct stages (bottom panel of Fig.~\ref{fig:kaitfig}, and
Fig.~\ref{fig:kcloseup}). During the first rise, the system became
much bluer in both $U-B$ 
and $B-V$.  During the second rise, it became bluer still in $B-V$,
but maintained an approximately constant
$U-B$ color.
On the final rise to optical maximum, the $U-B$ color again became
significantly bluer, suggesting a large increase in the temperature
and UV luminosity of the hot component,
whereas the $B-V$ color changed very little.

The optical decline was marked by 0.2--0.3 mag variations with no clear
pattern. The $U-B$ color remained quite blue and fairly constant for
roughly the first 300 days of the decline, and then began to redden.
The $B-V$ color reddened more gradually throughout the entire decline.
There was a small rebrightening around JD 2452600 (at the end of
2002), where the $U-B$ color initially reddened slightly, and then
became more blue.  However,
the $U-B$ color never became as blue as when the optical flux was at a
similar level during the main outburst.  The peak of the rebrightening
occured approximately 700 days after the main outburst peak.  The time
between the two peaks was thus slightly shorter than the orbital
period of 759 d
\citep{formleib94,mk96}, and similar to that of oscillations during
the decline of previous large outbursts \citep[e.g.,][]{kw84}. 

\subsection{Optical Spectroscopy} \label{sec:obs_optspec}

P. Berlind, M. Calkins, and several other observers acquired
low-resolution optical spectra of Z And with FAST \citep{fab98}, a
high-throughput slit spectrograph mounted at the Fred L. Whipple
Observatory 1.5-m telescope on Mount Hopkins, Arizona.  They used a
300 g mm$^{-1}$ grating blazed at 4750 \AA, a 3\arcsec-wide slit, and
a thinned Loral 512 $\times$ 2688 pixel CCD.  These spectra cover
3800--7500
\AA~at a resolution of 6 \AA.  
Starting in 1994, we obtained over 800 spectra of Z And, with over 350
during the 2000--2002 outburst.
Several additional spectra were obtained by 
R. Hynes 
near the optical peak with the 4-m William Herschel Telescope (WHT).
We list the dates of selected optical
spectroscopic observations 
taken near the beginning of the 2000--2002
outburst in Table~\ref{tab:optspec}.

We
reduced the spectra with IRAF.  After
trimming the CCD frames at each end of the slit, we corrected for the
bias level, flat-fielded each frame, applied an illumination correction,
and derived a full wavelength solution from calibration lamps acquired
immediately after each exposure.  The wavelength solution for each
frame has a probable error of $\pm$0.5 \AA~or better.  To construct
final 1-D spectra, we extracted object and sky spectra using the optimal
extraction algorithm APEXTRACT within IRAF.  Most of the resulting
spectra have 
S/N $\gtrsim$ 30 per pixel.

Optical spectra of Z And, the prototypical symbiotic star, generally
have strong high-ionization emission lines superimposed on a red-giant
absorption spectrum.  Our spectra show two types of long-term
variations. During quiescence, high-ionization lines, such as [\ion{Fe}{7}]
and the Raman-scattered \ion{O}{6} lines at 6830~\AA\, and 7088~\AA, are
roughly constant with orbital phase. 
Lower-ionization lines, including
the \ion{H}{1} Balmer lines, decrease significantly in intensity near phase
0.5, 
when the red giant lies in front of the hot component and obscures
part of the nebula.  

\begin{figure*}[t]
\begin{center}
\epsfig{file=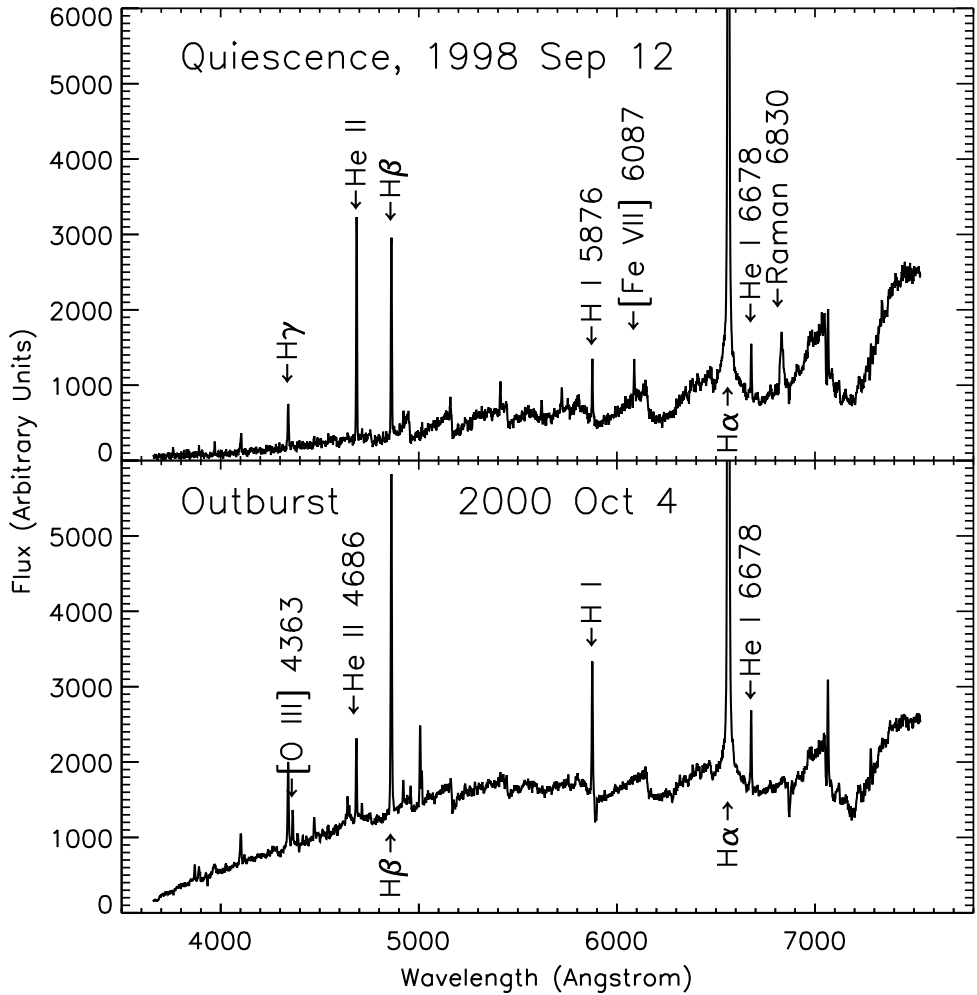,width=5.5in}
\end{center}
\caption{Examples of our Z And optical spectra from quiescence ({\it top}) and
outburst ({\it bottom}), taken with the FAST spectrograph at
Mt. Hopkins.  These two spectra were taken at similar orbital phase
(0.07 for the quiescent spectrum and 0.06 for the outburst spectrum,
according to the orbital ephemeris of
\cite{mk96}). \label{fig:qualspec}}
\end{figure*}

During outburst, all of the lines react to the
changing luminosity of the hot component as well as the orbital
geometry. In the 1997 outburst, the high-ionization lines increased in
intensity relative to the continuum, suggesting an increase in the
effective temperature of the hot component. The relative intensities
of these lines faded during a portion of the 2000 outburst,
implying a decline in the effective temperature of the hot component
(see \S\ref{sec:tempev}). Although we do not have high-resolution
spectra from the 1997 outburst, the WHT spectra taken near optical
maximum in late 2000 (December 2 and 3) show P-Cygni profiles for
the 
He I lines, consistent with spectra acquired during
previous major outbursts \citep[e.g.,][]{swingsstruve1970}.  At around
the same 
time, FUV spectra 
also show such profiles (see
\S\ref{sec:obs_fuv}).  In Fig.~\ref{fig:qualspec}, we show two example
spectra --- one from quiescence, and one near the peak of the 2000
outburst --- both taken just after orbital phase zero.  We list the
equivalent widths of 12 lines in 
826 spectra taken between 1994 September 13
and 2003 December 29 in Table~\ref{tab:optews}, which is
available electronically.

\subsection{Far Ultraviolet} \label{sec:obs_fuv}

Nine observations of Z And (plus one short ``safety snap'') were
performed
between 2000 November 16 and 2003 August 4 with the Far Ultraviolet
Spectroscopic Explorer ($FUSE$) satellite;
see Table~\ref{tab:fuv}.  The $FUSE$ spectrographic
instrument and on-orbit performance are described by \cite{moos2000}
and
\cite{sahnow2000}.  The spectral format consists of eight channels fed
by four co-aligned prime-focus telescopes and Rowland-circle grating
combinations, with a separate slit for each telescope.  In four of
the eight channels, LiF overcoated optics are used to cover the bandpass
from $\sim 1000$ to 1190~\AA.  The other four channels use SiC
overcoated optics to cover the bandpass from the Lyman edge to
$\sim 1100$~\AA.  
Each of the nine observations, consisting of anywhere from one to eight
orbits, was taken through the low wavelength resolution slit (LWRS),
which is 30~$\times$30~ square arcseconds in angular area.  

We reduced the $FUSE$ data with CALFUSE pipeline versions 2.0.5 and
higher, and combined the data for each orbit without considering
shifts in the stellar position within the slits during the
Earth-occultation period between orbital acquisitions.  The individual
spectral channels were linearized to a 0.01~\AA\ scale.  They were
then joined into one spectrum by cross correlation of overlapping
spectral sections.

For each of the wavelengths listed in Table~\ref{tab:fuv}, we
calculated the continuum level from the unweighted mean of all data
points at the specified wavelengths $\pm 0.2$~\AA,
from all the individual exposures at a particular epoch.
We took as the uncertainty the larger of the formal
statistical error in that mean 
and the scatter of the measurements from the individual exposures.
In a few cases, where one or two of the individual exposures at a given
epoch were discordant (centering in the aperture may have been poor
for several orbits in those cases), we removed those exposures.

\begin{figure*}
\vspace{-0.3cm}
\begin{center}
\epsfig{file=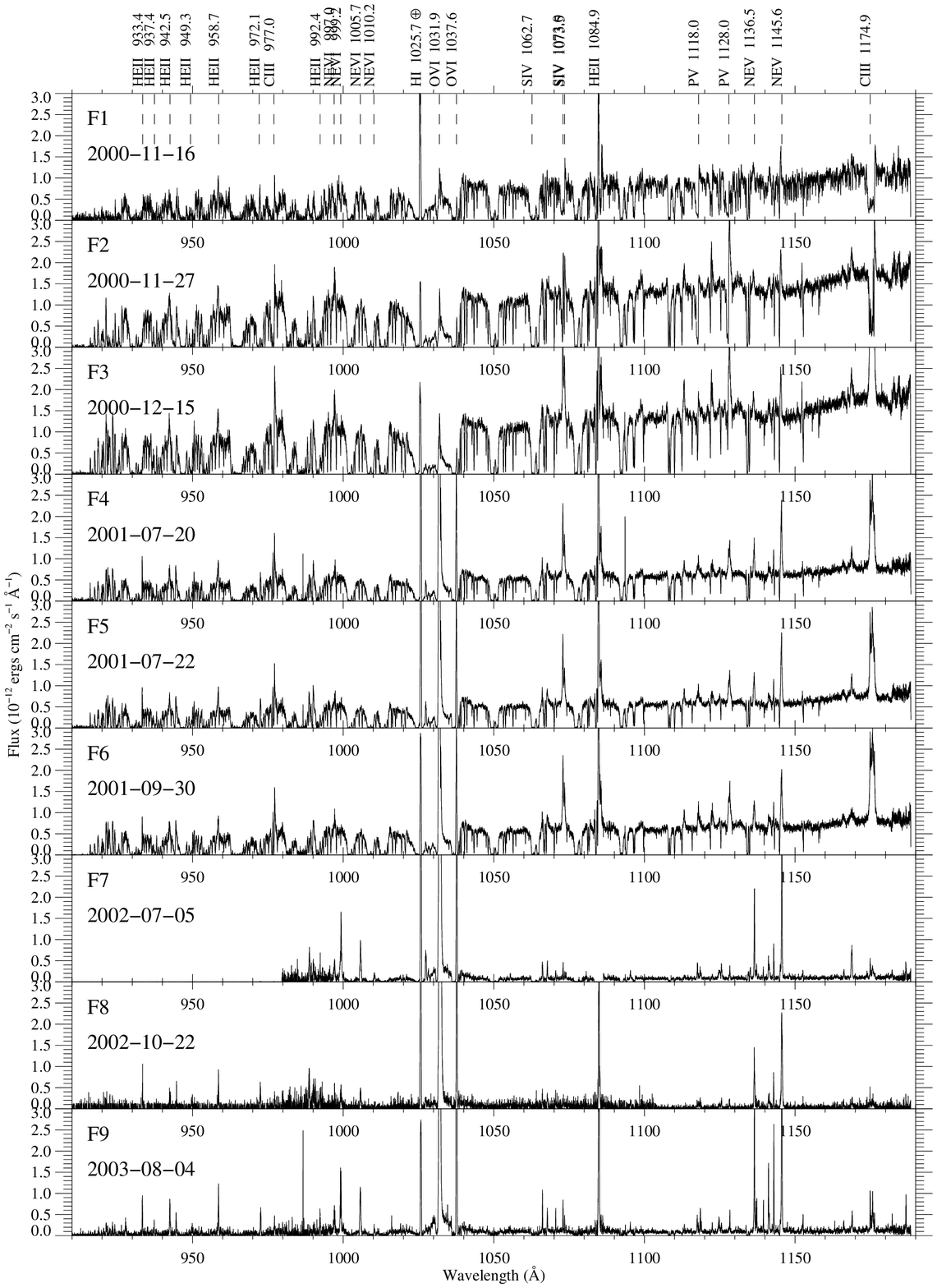,width=6.7in}
\end{center}
\caption{$FUSE$ spectra during and after
the 2000--2002 outburst.
The continuum level rises and falls, and lines such as \ion{O}{6}
and \ion{P}{5}
move from absorption to 
emission. 
\label{fig:fusespec}}
\end{figure*}

\begin{figure}[t]
\includegraphics[scale=0.47]{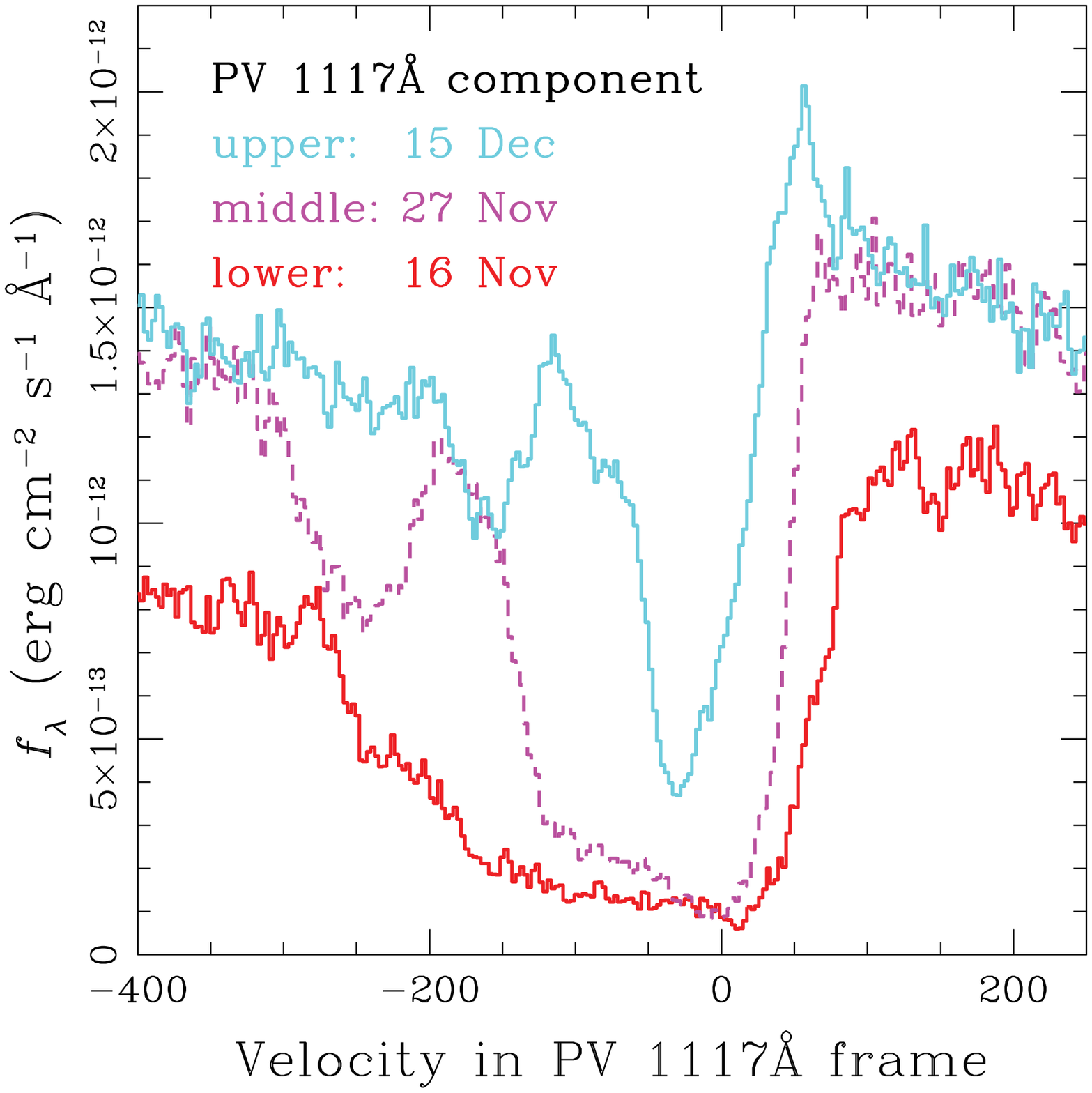}
\hspace{0.5cm}
\includegraphics[scale=0.47]{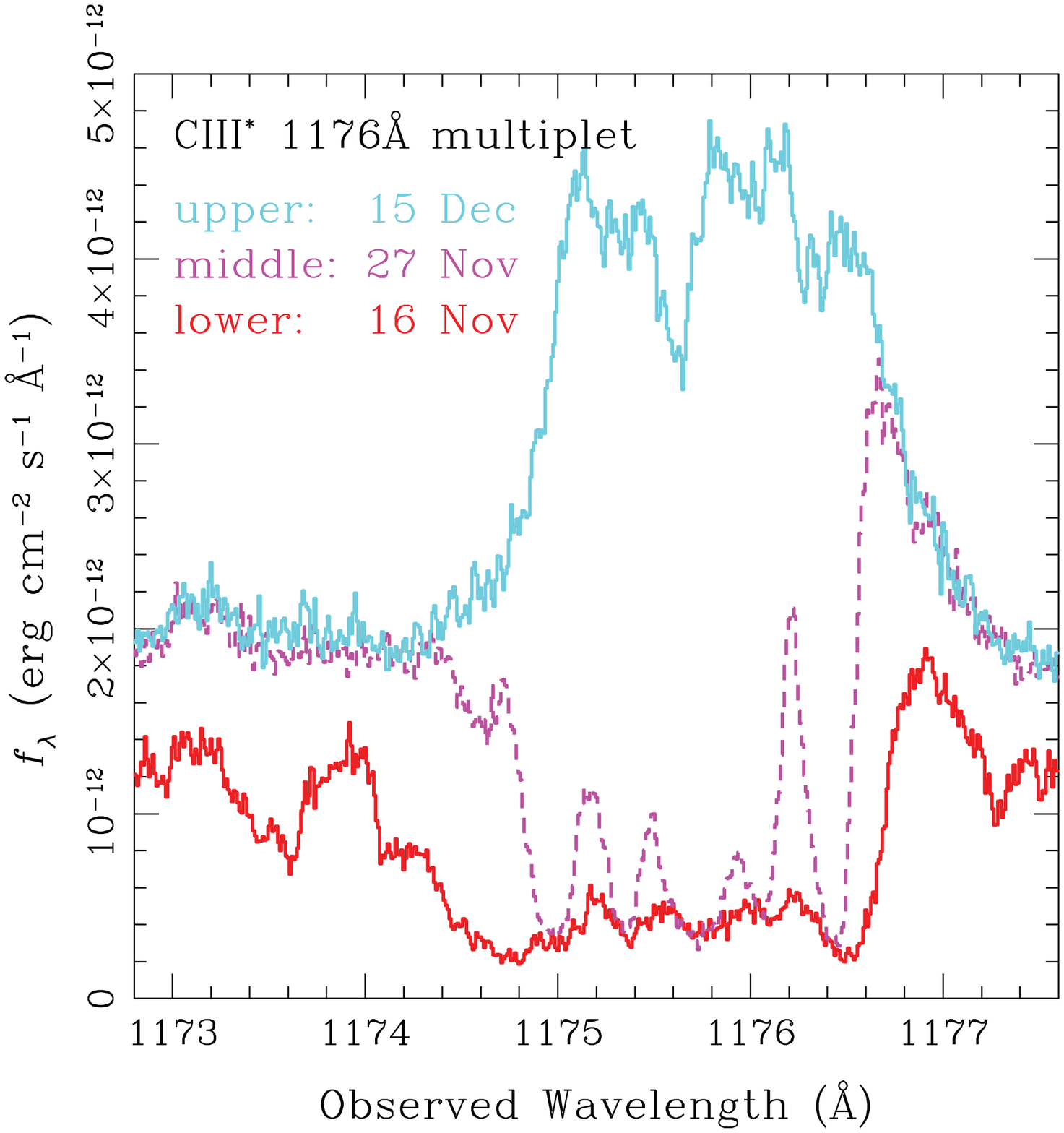}
\caption{Top panel: \ion{P}{5} ($\lambda$1117)
line profile from the first three $FUSE$ observations, in late 2000.
The absorption trough becomes narrower and the emission component
strengthens as the outburst progresses. Bottom panel: \ion{C}{3}
($\lambda$1176
multiplet) line profile from the same three $FUSE$
observations. The line complex evolves from absorption to
emission. \label{fig:fusecloseup}}
\end{figure}

Fig.~\ref{fig:fusespec} shows the joined $FUSE$ spectra (except for the
safety snap).  
When the object may have shifted out of the slit, we adjusted the flux
level to reflect the reduced integration time. 
The spectral continua are
rather flat, and strong absorption by interstellar 
H$_2$ spans the entire 
wavelength region
\cite[see][for H$_2$ absorption templates]{mccan03}.
During the main optical outburst, the FUV continuum rose and then fell,
peaking 
at around the same time as the optical maximum.  The continuum flux
variations are also evident from the 959.5~\AA, 1058.7~\AA, and
1103.4~\AA\, flux densities listed in the last three columns of
Table~\ref{tab:fuv}.  By the seventh $FUSE$ observation (F7), on 2002
July 5, the FUV flux had returned to approximately the quiescent value
recorded by $Orpheus$ and $HUT$ in 1993 and 1995, respectively
\citep[$Orpheus$ and $HUT$ data were obtained from the Multimission
Archive at STScI database, plus see][]{birrielea98,schmidea99}.
During the subsequent brief optical rebrightening in 2002, the FUV
flux dropped by almost an order of magnitude, and then rose back to the
quiescent level again in the final $FUSE$ observation on 2003 August 4.

In addition to absorption by some high-ionization species, the
first $FUSE$ observation shows a large amount of cool gas shrouding
the FUV continuum.  We see singly and doubly ionized C,
Fe, and Si, as well as absorption by \ion{P}{5} and the non-resonance
1176~\AA\, transition in \ion{C}{3}. 
Low-ionization species with P-Cygni profiles include \ion{N}{2},
\ion{P}{2}, \ion{C}{2}, and
\ion{Fe}{3}.  The presence of P-Cygni profiles
in $FUSE$ data is consistent with the optical P-Cygni
profiles at around the same epoch.
By the time of the second $FUSE$ observation, the absorbing material begins to
clear away.  Moreover, whereas the first spectrum is almost entirely
in absorption, the later spectra contain increasingly strong emission
features.
Toward the end of the series of FUV spectra, \ion{O}{6} becomes extremely strong as
the continuum level falls.

Fig.~\ref{fig:fusecloseup} shows two examples of 
the early evolution of some line profiles.
\ion{P}{5} begins in absorption, with a blueshifted (P-Cygni-type)
line width of approximately 300 km s$^{-1}$.  In the second $FUSE$
observation, taken immediately before optical maximum, the width of
the main blueshifted \ion{P}{5} absorption decreases to less than 200
km s$^{-1}$, and by the third $FUSE$ observation, taken just after
optical maximum, the \ion{P}{5} absorption components have narrowed
further and an emission component is present.
During this same time, \ion{C}{3}
$\lambda$1176
moves from absorption to emission.

The \ion{P}{5} absorption-line ratio also demonstrates the 
trend of decreasing opacity evident in the line profiles from the first
three $FUSE$ spectra.  For optically thin gas, the ratio of the
$\lambda$1117
to $\lambda$1128
components of \ion{P}{5} is 2:1, whereas for optically thick gas, the
ratio is 1:1.  In the first two $FUSE$ observations, the ratio of
the two \ion{P}{5} line components is close to 1:1.
By the third observation, however, the ratio is approximately 1.33:1,
indicating that the opacity had begun to decrease.
Furthermore, although the early \ion{P}{5} line profiles suggest high
optical depth, the troughs of the lines do not touch zero.  Either 
the $\tau = 1$ surface probed in
the 
absorption line has significant emissivity relative to the underlying
continuum, another continuum source is present, or the optically thick
material only partially covers the continuum source\footnote{Nebular
recombination radiation (predominantly from \ion{He}{2}) is unlikely
to be the possible secondary source of FUV emission since
the unabsorbed flux
expected from the nebula at FUV wavelengths is only a few
times $10^{-13}\, {\rm ergs\, cm}^{-2}\, {\rm s}^{-1}$
\citep{fc88, fc95}.  After extinction by the interstellar medium,
the nebula contributes on the order of $10^{-14}\, {\rm ergs\,
cm}^{-2}\, {\rm s}^{-1}$ at these wavelengths, which is much less than
the flux we see below the optically thick line profiles.
}. The early
\ion{C}{3} and \ion{Si}{3} lines also appear to be optically
thick, but do not reach zero.

\subsection{Radio} \label{sec:obs_radio}

\begin{figure*}[t]
\begin{center}
\epsfig{file=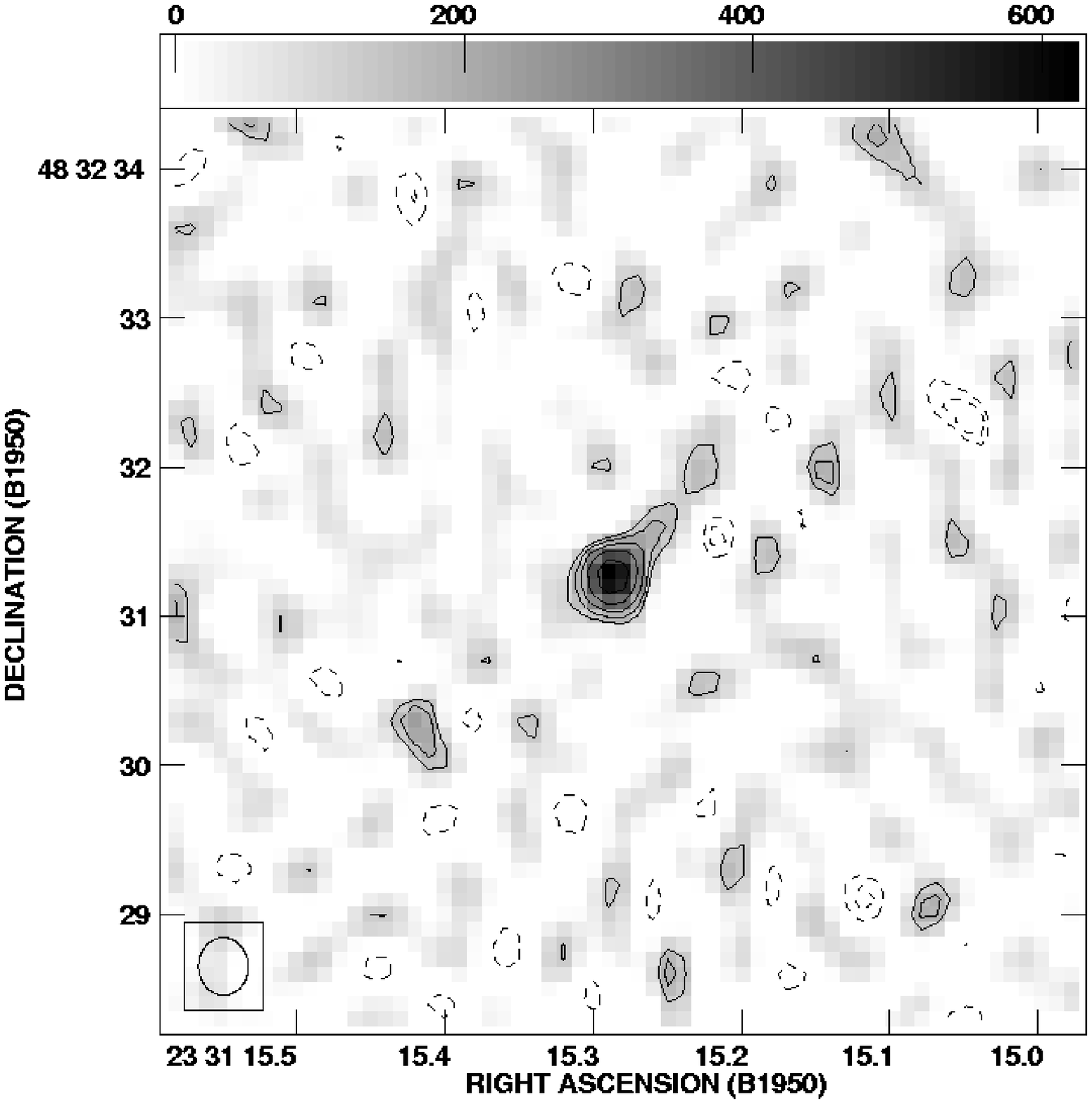,width=5in} 
\end{center}
\caption{Extended 5 GHz radio emission detected on 2003 July 24 with the VLA
(in the A configuration).  The extension is in a direction roughly
20$^\circ$ west of the small, transient ejection seen in 2001 September,
just one year after the beginning of the optical outburst, by
\cite{brock04}.   The contour levels are $6.1 \times 10^{-2}\, \times$
($-4$, $-2.83$, $-2$, 2, 2.83, 4, 5.657, 8) mJy.
\label{fig:radiojet}}
\end{figure*}

Radio observations were performed with the NRAO Very Large Array (VLA)
in New Mexico, and the Multi-Element Radio Linked Interferometer
Network (MERLIN) in the U.K.  Reduction was done with AIPS, and
details of the reduction and analysis of the first 12 observations
(R1-R12) are given in \cite{brock04}.  For the final three
observations (R13-R15; from the VLA),
we observed in standard continuum mode, with two
intermediate-frequency (IF) pairs of sidebands 
recording both right and left circular polarization, for a total
bandwidth of 200 MHz at each frequency (100 MHz per polarization for
two polarizations). 

For phase calibration, we used a 110 s / 50 s cycle at 8.5 GHz and a
100 s / 50 s cycle at 5 GHz.  The phase calibrator, 2322+509, is
2.8$^\circ$ away from Z And.  Flux densities are defined with respect
to 0542+498 (3C 147) and 0137+331 (3C 48), using the 1999.2 VLA
flux-density scale.
The flux densities for Z And
represent the average of the two IF sideband pairs at each band,
integrated 
over a small region surrounding the peak.  
We constructed
images of the full primary beam at each frequency, using natural
weighting for maximum sensitivity.
These images were deconvolved with the CLEAN algorithm to remove the
sidelobes of both Z And and (at 4.9 GHz) a nearby, extended confusing
source.
We list radio flux densities between 2000 and 2003 in
Table~\ref{tab:radio}.
The 5 GHz flux densities 
are also plotted in the bottom panel 
of Fig.~\ref{fig:tempev}.  

Early in the optical outburst, the 5 GHz flux density was very low
compared to typical quiescent levels of about 1 mJy.
Our first radio data were obtained
approximately 40 days after the beginning of the optical rise, so 
the radio flux density could have dropped before the initial optical
rise, simultaneous with the rise, or after the rise.  As the optical
outburst progressed, Z And brightened in the radio and, at 15 GHz,
reached $3.5 \pm 0.1$ mJy, the highest 15 GHz flux density ever
recorded for this source.
The radio brightness peaked between two and five months after the
optical brightness maximum (first at 5 GHz, then at 15 GHz, and finally
at 1.4 GHz).  During the optical rebrightening in late 2002,
the radio brightness temporarily dropped below typical quiescent
levels.

Roughly one year after the start of the optical outburst, and about
eight and six months after the respective 5 GHz and 15 GHz brightness
maxima,
\cite{brock04} discovered a radio jet from Z And.  The extended radio
component was just 60 mas in length, but was oriented in a direction
consistent with the outflow being almost exactly perpendicular to the
binary orbital plane
\citep[see][]{schsch97}.  In 2001 September, the ejected component was brighter
than the central component.  However, the extended component was not
detected in three subsequent MERLIN observations taken over the course
of the next eight months.  Z And was unresolved by the VLA (in C
configuration) in 2002 December and 2003 January.

In 2003 July, our VLA observations (in A configuration) indicated that
the radio jet was again detectable, and had reached $\sim$0.5$\arcsec$
to $\sim1\arcsec$ in length.  A Gaussian fit to the 5 GHz data
suggests that the source was extended, with a nominal major-axis
full-width at half maximum (FWHM) of $0.32\arcsec \pm 0.08\arcsec$ and
a minor axis less than 0.15\arcsec, at an orientation of $130^\circ
\pm 15^\circ$ east of north.  For comparison, the beam is a Gaussian
with FWHM $0.385\arcsec \times 0.333\arcsec$ oriented 1.6$^\circ$
degrees east of north.  In the map, the source shows an elongation to
the NW (see Fig.~\ref{fig:radiojet}), and we obtained similar results
with different $uv$-weighting schemes.
The peak flux density of the extended component in the 5 GHz VLA image is
about 0.18 mJy beam$^{-1}$.  There is no sign of extended emission in the
15 GHz image.  The 2003 July VLA extension is about 20$^\circ$ west of
the direction of the 2001 September MERLIN jet, but in a similar direction
as a blob in a 1991--1992 VLA+MERLIN map shown by \cite{kennyphd}.  

If we take the size of the VLA extended region at face value and
assume that the material was ejected at the start of the optical
outburst, the outflow has a velocity of around 800 to 1000 km
s$^{-1}$.  The inferred velocity of the small (60 mas) jet found with
MERLIN was 400 km s$^{-1}$.  However, the two velocity estimates have
sufficiently large uncertainties that they are roughly consistent.

\subsection{Fast Optical Photometry} \label{sec:obs_rapidvar}

On 2000 November 23,
we performed fast optical photometry using the 1-m Nickel telescope at
UCO/Lick Observatory.  The goal of these observations was to determine
whether the 28-minute oscillation due to accretion onto the spinning,
magnetic WD was detectable during the 2000--2002 outburst, as it was
during the event in 1997 \citep{sb99}. We took repeated 15-s exposures
using a $B$ filter and an
unthinned, phosphor-coated, 2048~$\times$~2048 pixel Loral CCD.
A typical 
dead-time (with 4$\times$4
binning and fast readout) of 23 s 
gave a time resolution of 38 s.
The full time series spanned 5.6 hours.
A description of our differential photometry for this field and the data
reduction and analysis methods can be found in
\cite{sb99} and \cite{sbh01}, respectively.

The 28-minute oscillation was not detected.  After removal of some
hour-time-scale variability with low-order polynomial fitting, we
found an upper limit on the amplitude of any 28-minute signal in the
data of 3 mmag \citep[see][for a description of how we obtain upper
limits using Monte Carlo simulations]{sbh01}.  The conservative upper
limit of 3 mmag was obtained without use of the brightest comparison
star, GSC 03645-01592, since this star is an intrinsically variable
$\delta$-Scuti star \citep{sbcf02}.  If GSC 03645-01592 is used, and
its intrinsic variability removed with a high-order polynomial, the
upper limit for the 28-minute oscillation becomes 1.8 mmag.  

This non-detection indicates that the evolution of the 28-minute
oscillation during the large 2000--2002 outburst was fundamentally
different from its evolution during the smaller 1997 outburst.  In
1997, the oscillation amplitude changed roughly in step with the
hot-component optical flux, reaching almost 6 mmag when the optical
flux was 0.9 mag above the pre-outburst level.  In 2000, despite the
fact that Z And had brightened by 1.8 mag compared to quiescence, the
oscillation was not detected.

\subsection{X-Ray} \label{sec:obs_xray}

We obtained three X-ray observations of Z And during the first year of
the 2000--2002 outburst.  The first observation (X1) was performed
with the $Chandra$ X-Ray Observatory \citep{weissk02} on 2000 November 13,
just over two months into the optical outburst, at the beginning of
the third stage of the optical rise.  The subsequent two X-ray observations
(X2 and X3) were performed with the {\it XMM-Newton} Observatory
\citep{aschenb02}, on 2001 January 28 ($\sim$50 d after the optical
outburst peak) and 2001 June~11 (6 months after the optical peak),
respectively.  See Table~\ref{tab:xray} for a summary of all three
X-ray observations.

Observation X1 was performed with the {\it Chandra} High Energy
Transmission Grating Spectrometer (HETGS).  The data were processed
using updated calibration files (CALDB 2.27), and reduced with the
software package CIAO v3.1.  In 19~ksec, the source was detected with
109 source counts in the zeroth-order spectrum.  Due to the low number
of counts, we used only the zeroth-order data for our analysis. To
reduce the instrumental background, we only selected photons with
energies between 0.3 and 7 keV.  Based on examination of source-free
regions in the S1 chip, no periods of high background occurred during
the observation.

We performed spectral analysis for all X-ray data using the software
packages Sherpa (within CIAO) and XSPEC v11.3.  For observation X1, we
extracted the source spectrum from a $3''$-radius circular region
centered on the source, and used an annular region also centered on
the source for the background. We rebinned the spectrum to have at
least 10 counts per spectral bin.  In the energy range over which the
source was detected, the background contributed less than 0.1\%.

Observations X2 and X3 were performed with the {\it XMM-Newton}
European Photon Imaging Cameras (EPIC), in full-frame mode, using the
medium neutral-density filter.  After rejecting data with high
background levels, we retained good-time intervals of 16~ksec and
4~ksec (out of 20~ksec and 15~ksec) for observations X2 and X3,
respectively.

\begin{figure*}[t]
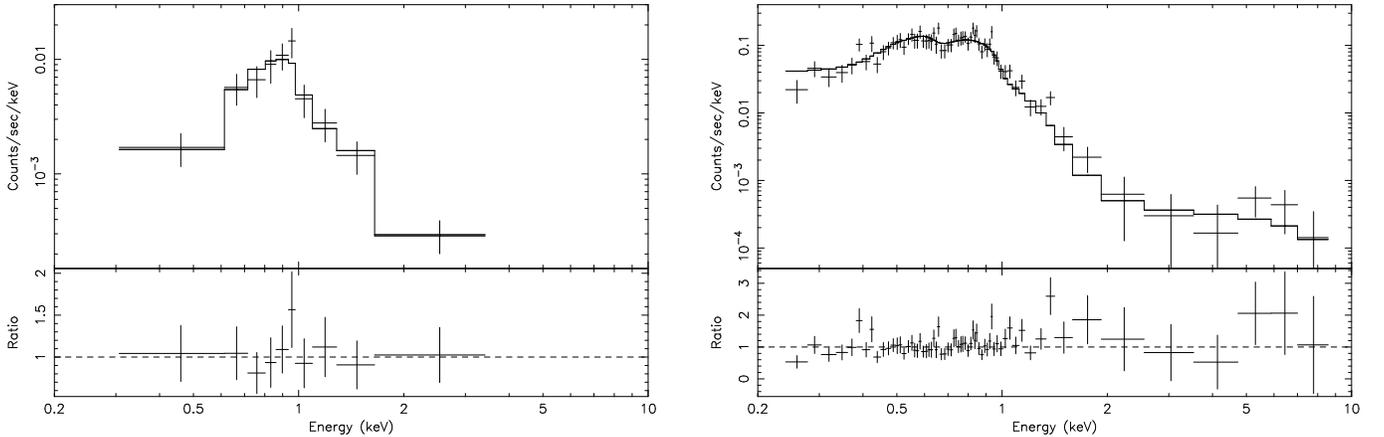

\includegraphics[angle=-90,scale=0.37]{f8a.eps}
\hspace{0.5cm}
\includegraphics[angle=-90,scale=0.37]{f8b.eps}
\caption{Left: X-ray spectrum of Z~And from the 2000 November {\it
Chandra} observation (X1), fitted with an absorbed blackbody model
($kT=0.2$~keV) plus an absorption edge at $\sim 1$~keV. Right: X-ray
spectrum of Z~And from the 2001 January {\it XMM-Newton} observation (X2),
fitted with an absorbed blackbody ($kT=0.11$~keV) and powerlaw
($\alpha=0.79$) model with absorption edges at 0.29~keV (most likely due
to a calibration error), 0.64~keV, and 0.96~keV. \label{fig:xrayspecs}}
\end{figure*}

We reduced and analyzed the {\it XMM-Newton} data with the software
package SAS v5.4.1, using only data in the 0.2--10 keV range.  We
extracted the EPIC-pn source spectra from a $30''$-radius circle
centered on the source, and the background from a source-free region.
Z~And was detected in the EPIC-pn with over 1500 source counts during
observation X2, but was significantly fainter (about 140 EPIC-pn
source counts) during observation X3.  In addition, the background
level was significantly higher during observation X3 (approximately
0.7 counts s$^{-1}$, probably due to solar activity, compared to $<
0.05$ counts s$^{-1}$ in observation X2).  To improve the statistics,
we rebinned the spectra from X2 and X3 to have 20 and 10 counts per
spectral bin, respectively.

The X-ray emission detected in observation X1 fell primarily between
0.3 and 2 keV. With only $\sim$100 source counts and a large
uncertainty in the amount of intrinsic absorption, we are unable to
use the spectrum from X1 to strongly constrain physical models.
Nevertheless, we fit the spectrum with several different models as a
means of parametrizing its softness and investigating the absorption.
We could not obtain an acceptable fit for a single-component
Raymond-Smith (RS) plasma model.  For a powerlaw model, the best fit
is statistically acceptable, but gives an unphysically large photon
index of $\Gamma = 5.9$ (where the photon flux is given by $A(E) = K
E^{-\Gamma}$, with $E$ being energy and $K$ a normalization constant).
The absorbing column density in the powerlaw model is $N_H =
5\times10^{21}$ cm$^{-2}$, which is significantly higher than the
Galactic value of $1.4\times10^{21}$ cm$^{-2}$ \citep{dl90}, but lower
than values of $\sim10^{22}$ cm$^{-2}$ sometimes found for symbiotic
stars near orbital phase zero
\citep[where X-ray emission from the accreting white dwarf must pass
through the maximum amount of the red-giant wind;][]{vogel91}.  As
discussed in \S\ref{sec:lum}, we adopt $E(B-V) = 0.27$~mag for Z And.
Based on the relationship between $E(B-V)$ and $N_H$ shown in Fig.~1.1
of \cite{spitzer}, $E(B-V)=0.27$~mag is consistent with the Galactic
column density given by \cite{dl90}.

A blackbody model yields a good fit if we include an absorption edge
at $\sim 1$ keV; the significance level of the additional absorption
edge is greater than 95\%.  The resultant blackbody temperature is
0.21$\pm 0.04$ keV, or $T_{bb} = (2.4 \pm 0.5) \times 10^{6}$~K, and
in this model $N_H$ is consistent with the Galactic value.  The
absorbed flux in the energy bands 0.3--7 keV and 2--20 keV is $(2.5\pm
0.2)\times10^{-13}$ erg cm$^{-2}$ s$^{-1}$ and $(6.9 \pm 0.7)
\times10^{-15}$ erg cm$^{-2}$ s$^{-1}$, respectively.  The spectrum
from observation X1, with the blackbody-plus-absorption-edge model, is
shown in the left-hand panel of Fig.~\ref{fig:xrayspecs}.  We list the
best-fitting spectral parameters in Table~\ref{tab:xfits}.

For observation X2, there is no single-component spectral model which
can provide an acceptable fit to the data, in part because low-level
emission was detected all the way out to 10 keV.  As with observation
X1, a single-component RS plasma model does not generate a good fit.
Instead, to obtain a reasonable $\chi^2_{\nu}$, a blackbody component
with $kT = 0.11^{+0.02}_{-0.01}$ keV ($T_{bb} = (1.3 \pm 0.2) \times
10^6$~K), a powerlaw component with $\Gamma = 0.79$, and two
absorption edges (at 0.64 and 0.96~keV) are required.  The optical
depth of both edges is $\sim$1.  An additional edge at 0.29 keV is
probably due to the calibration error near the carbon edge at 0.3~keV.
The derived $N_H$ in this model is consistent with the Galactic value.
The powerlaw component comprises about 6\% of the total unabsorbed
flux.  The absorbed flux in 0.3--7 keV and 2--20 keV is
$(1.3\pm0.04)\times10^{-13}$ erg cm$^{-2}$ s$^{-1}$ and $(6.8\pm
0.2)\times10^{-14}$ erg cm$^{-2}$ s$^{-1}$, respectively. The spectrum
from observation X2 and best-fitting spectral model are shown in the
right panel of Fig.~\ref{fig:xrayspecs}.  The spectral-fit parameters
are listed in Table~\ref{tab:xfits}.

Z~And was significantly fainter during observation X3.  With the low
source count rate and high background level during this observation,
it is clear that the spectrum was soft, but it is otherwise not
tightly constrained.  The spectrum is best described by an absorbed
blackbody model with $kT=0.12\pm0.07$~keV ($T_{bb} = (1.4 \pm 0.8)
\times 10^6$ K), and $N_H = 2.4^{+4.8}_{-2.4} \times 10^{21}\, {\rm
cm}^{-2}$ A powerlaw with an unphysically high photon index of $> 10$
produced a similar-quality fit, as did a RS plasma model fit, which
gave $kT = 0.086\pm 0.002$ keV ($T_{RS} = (1.00\pm 0.02) \times 10^6$
K) and an absorbing column close to $10^{22}\, {\rm cm}^{-2}$ ($N_H =
(9.5 \pm 0.05) \times 10^{21}\, {\rm cm}^{-2}$).  The absorbed fluxes
in 0.3--7 keV and 2--20 keV are $(2.0 \pm 0.5)\times10^{-14}$ erg
cm$^{-2}$ s$^{-1}$ and $(6.1\pm2)\times10^{-18}$ erg cm$^{-2}$
s$^{-1}$, respectively.

In all three X-ray observations, we find formally acceptable fits to
blackbody models with temperatures of a few times $10^6$ K.  However,
these temperatures are too high for the X-ray emission to be blackbody
emission from either the WD surface or accretion disk.  A WD with a
radius of $10^9$ cm and blackbody temperature $T_{bb} = 2
\times 10^6$ K would have a luminosity of $3
\times 10^6 \lsun$ --- two orders of magnitude greater than the Eddington
limit for a $0.65\, \msun$ WD.  Moreover, our optical and FUV data
indicate that the effective temperature of the WD in Z And typically
does not rise above 180,000 K (see
\S\ref{sec:templumev}). Given that we have very poor statistics for
two of the three X-ray observations, that symbiotics generally have
rich line spectra at other wavelengths, and that there are multiple
potential sites of X-ray emission and absorption in Z And, we suspect
that a more complex X-ray spectrum, possibly containing overlapping
emission lines from a plasma and absorption features from the ionized
nebula, is mimicking a blackbody in our CCD-resolution spectra.

In both the FUV (see \S\ref{sec:obs_fuv}) and radio (see
\S\ref{sec:obs_radio}) data, we find evidence that a significant
amount of material was ejected from the WD into the dense symbiotic
nebula.  Thus, an alternative possible source of X-ray emission from Z
And during outburst is shock-heated plasma.  Taking a distance of 1.2
kpc, the unabsorbed X-ray fluxes from Table~\ref{tab:xfits} give
0.3-7.0-keV X-ray luminosities of almost $10^{32}$ erg s$^{-1}$ for
observation X1, dropping to a few times $10^{31}$ erg s$^{-1}$ by
observation X3.  These X-ray luminosities are
more than five orders of magnitude less than the total bolometric
luminosity of the WD near the peak of the outburst (see
\S\ref{sec:lum}).  Thus, most of the outburst energy is emitted at slightly
lower than X-ray energies, as expected given the WD effective
temperature of $< 180,000$~K.
The total 0.3-7-keV X-ray energy radiated in four months (the
approximate time for which the X-ray emission was high compared to
quiescence) is about $8 \times 10^{38}$~erg --- less than the
kinetic energy of $10^{-9}
\msun$ of material moving at 300 km s$^{-1}$ (the approximate velocity
of the outflowing material; see \S~\ref{sec:obs_fuv} and
\S~\ref{sec:obs_radio}).  Since at least $10^{-7} \msun$ was probably accreted
in 2000 (see \S\ref{sec:implications}), the availability of this much
kinetic energy for conversion to X-ray emission is plausible.  In
addition, the temperature behind a strong shock moving at 400 km
s$^{-1}$ is roughly $2 \times 10^{6}$ K ($10^{6}$ K if the shock
velocity is 300 km s$^{-1}$), and plasma at this temperature will
produce soft X-ray emission.  Moreover, if we take a cooling time of a
few months and a plasma temperature of $10^6$ K, we infer a post-shock
density on the order of $10^7$ cm$^{-3}$.  This estimate gives a
pre-shock density of a few times $10^6\, {\rm cm}^{-3}$, which is
reasonable for the outer nebula in Z And \citep[assuming that the
density decreases from the inner-nebula value of $\sim10^{10}\, {\rm
cm}^{-3}$, from][as the inverse of the distance squared]{fc88}.

The presence of absorption edges provides additional evidence for
significant amounts of hot, possibly shock-heated gas.  In observation
X1, we found a 1.02~keV edge that is close to the energies of either
the Fe VII L$_1$ edge at 1.03~keV or the Ne VI K edge at 1.04~keV.
\cite{ebisawa01} saw the same edge in observations of the supersoft
X-ray source RX~J0925.7--4758, and they ascribed it to a high column
of ionized material between the WD and the observer.  We also detected
a 0.96~keV edge in the observation-X2 spectrum, possibly related to
the same feature.  The 0.64~keV edge seen in the observation-X2
spectrum is close to the Lyman edge of \ion{N}{7} at 0.67~keV. A
similar feature was found in the X-ray spectrum of the recurrent nova
U Sco
\citep{kahabka99}.  
Since it is hard to imagine that the hot-component effective
temperature could get high enough to produce \ion{N}{7} via
photoionization in Z And, the presence of the 0.64~keV edge could be
evidence for shock heating.  The hard tail in observation X2 provides
additional evidence for shock-heated gas at this time.

Comparing our X-ray fluxes to those from a 1993 quiescent-state
$ROSAT$ PSPC observation
\citep{murset97}, 
the X-ray flux of Z And increased by about an order of magnitude
during the 2000--2002 outburst.  The PSPC count rate in the $ROSAT$
energy band (0.12--2.4~keV) in 1993 was $0.003 \pm 0.002$ counts
s$^{-1}$, equal to approximately $3 \times 10^{-14}\, {\rm erg}\, {\rm
cm}^{-2}\, {\rm s}^{-1}$ (for reasonable blackbody spectral models).
In observations X1 and X2, the 0.3--2.0 keV flux was at least
$10^{-13}\, {\rm erg}\, {\rm cm}^{-2}\, {\rm s}^{-1}$ or a few times
$10^{-13}\, {\rm erg}\, {\rm cm}^{-2}\, {\rm s}^{-1}$.  By the time of
observation X3, six months into the optical decline, the X-ray flux
had decreased to a level similar to that measured by $ROSAT$ during
quiescence.

We also examined two public Rossi X-ray Timing Explorer PCA
\citep{jahodaea96} observations of Z And --- one during the decline of
the small optical outburst in 1997 (45~ksec; 1997 November 28), and
another a year later after the system had returned to quiescence
(12~ksec and 37~ksec on 1998 November 17 and 18).
In the first $RXTE$ observation, Z And was
detected with a rate of $\sim$2 counts s$^{-1}$.
Using the software PIMMS 
and assuming a powerlaw spectral model, this count rate gives a 2--20
keV flux of a few times $10^{-12}\, {\rm erg}\, {\rm cm}^{-2}\, {\rm
s}^{-1}$, two to three orders of magnitude higher than the 2--20 keV
fluxes estimated for the 2000--2002 outburst.  Z And was not detected
during the second, quiescent-state $RXTE$ observation.  Thus, unlike
in the supersoft X-ray sources, some observations of which show an
inverse relationship between the optical and X-ray brightness
\citep[][]{southwell96}, in Z And the X-ray flux rose during both the
1997 and 2000--2002 optical outbursts.

\section{Analysis and Conclusions} \label{sec:analysis}  

To investigate the changes in the hot component from 1994 to 2003, and
to diagnose the nature of the two outbursts during this
time period, we use the data described above to estimate the effective
temperature and luminosity of the WD-plus-disk as a function of time.
In addition, optical and FUV line profiles provide information about
material ejected during the 2000--2002 outburst.
 
\subsection{Hot-Component Effective-Temperature and Luminosity Evolution}
\label{sec:templumev} 

\subsubsection{Effective Temperature} \label{sec:tempev}

We estimated the effective temperature of the hot component
between 1994 and 2003 using both
the ratio of nebular emission lines \citep{iijima81} and the presence
of different ionization states in
the optical and FUV spectra \citep{mursnuss94}.  Assuming a
radiation-bounded nebula (all photons capable of ionizing H and He
ionize these atoms),
\cite{iijima81} showed that the effective temperature of a central source
of ionizing photons is 
\begin{equation} \label{eq:tiijima}
T_{Iijima} = 19.38 \left( \frac{2.22 F_{4686}}{4.16F_{H\beta}+9.94
F_{4471}} \right)^{1/2} + 5.13, 
\end{equation} 
where $F_{4686}$, $F_{H\beta}$, and $F_{4471}$ are the nebular
emission-line fluxes of \ion{He}{2}~$\lambda$4686, H$\beta$, and
\ion{He}{1}~$\lambda$4471, respectively.  This expression is
valid for effective temperatures between 70,000~K and 200,000~K. In
our treatment, we neglected the $\lambda$4471 flux, because F$_{4471}$
$\lesssim$ 0.1 F$_{H \beta}$.
To minimize calibration errors, we used the ratio of
equivalent widths, EW$_{4686}$/EW$_{H \beta}$, instead of the flux
ratio. Because the continuum is nearly flat at $\lambda
\lambda$4650--4900, this approximation is justified. 
Errors in the
equivalent widths are $\sim$ 5\%--10\%; thus, the statistical error in
the effective temperature is $\sim$ 10\%--15\%. 
To correct for the fact that near orbital phase zero the
H$\beta$-emitting region is partially occulted by the red giant, and
thus the H$\beta$ equivalent widths (EW$_{H \beta}$) do not reflect
the true H$\beta$ flux from the nebula at these times, we multiplied
$T_{Iijima}$ by a sinusoidal factor,
$T_{hot} = T_{Iijima}
(4/(5+\cos(2\pi\theta)))$, where $\theta$ is the orbital phase.  This
expression was empirically derived to 
force the quiescent orbital-phase-zero
values of $T_{hot}$ on average
to equal the quiescent orbital-phase-0.5 values.

We also estimated $T_{hot}$ based on the presence of different
ionization states in
the optical and FUV spectra.  
\cite{mursnuss94}  derive a relation between $T_{hot}$ (in degrees K) and
$\chi_{max}$, the ionization potential (in units of eV) of the
highest-ionization
species visible in the spectrum:
\begin{equation}
T_{hot} = 1000 \chi_{max}.
\end{equation}
They used photoionization models to estimate an accuracy of $\sim$10\%
for this expression.  
The ionization-state-based estimate of $T_{hot}$ for Z And generally
agreed with the line-ratio-based estimate described in the previous
paragraph.  However, the two methods gave slightly different results
at the time of maximum WD luminosity during the 2000--2002 outburst,
when the assumption of a radiation-bounded nebula might have been
violated (the rather low X-ray absorpting column and the
higher-than-average radio flux are both consistent with the nebula
being more ionized around the time of peak luminosity),
and near orbital phase zero in 2002, when the H$\beta$-emitting region
was partially occulted (as described above).  Because of these
problems with the line-ratio-based estimates, and because the
combination of optical and $FUSE$ spectra provides a good sampling of
ionization states that bracket the temperatures of interest, the
ionization-state-based results were given precedence on these
occasions.

\begin{figure*}[t]
\begin{center}
\epsfig{file=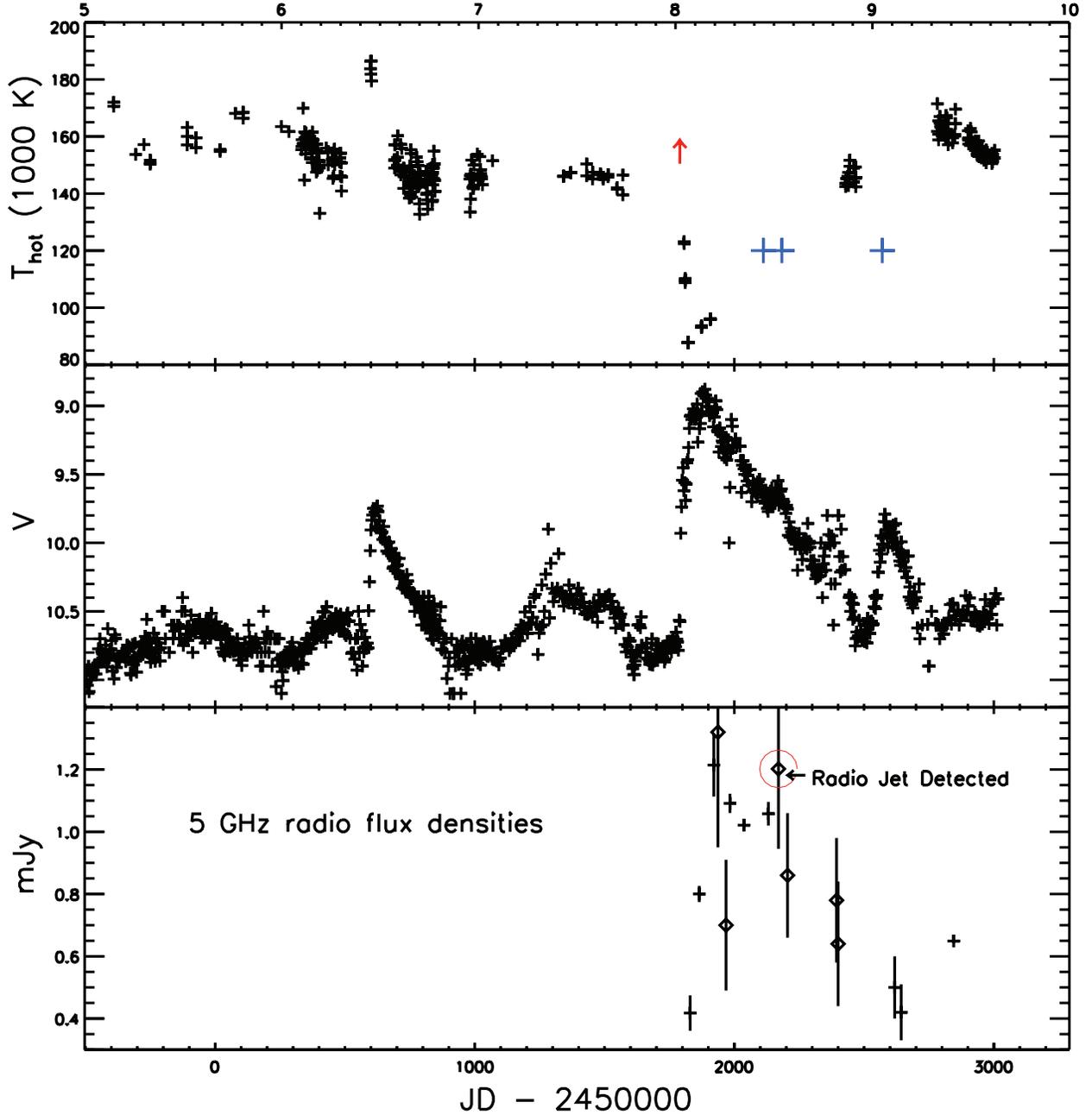,width=6.5in}
\end{center}
\caption{Top panel: the hot-component effective temperature,
$T_{hot}$, between 1994 and 2003.  The large blue crosses mark times
when our two methods for estimating $T_{hot}$ disagreed slightly, and
the ionization-state-based estimate was used (see text for details).
$T_{hot}$ rose during the 1997 outburst and dropped during the
2000--2002 outburst. $U$-band fluxes and $U-B$ colors near the very
beginning of the 2000-2002 outburst \citep[from][]{sko2003} indicate
that $T_{hot}$ probably rose briefly before dropping in 2000 (red
arrow).  Middle panel: $V$-band light curve, from the AAVSO, showing
both the 1997 and 2000--2002 outbursts.  Bottom panel: 5 GHz radio
flux densities, which drop (compared to the typical quiescent level of
about 1 mJy) early in the 2000--2002 outburst.  Orbital phase from the
ephemeris of \cite{mk96} is shown at the top of the plot.
\label{fig:tempev}}
\end{figure*}

For our spectra, the persistence of He II~$\lambda$4686 ($\chi =
54.4$~eV) indicates $T_{hot} \gtrsim$ 55,000 K throughout 1994--2003.
Both Raman-scattered \ion{O}{6} at $\lambda$6830 ($\chi = 114$~eV) and
the [\ion{Fe}{7}] $\lambda$6087 ($\chi = 100$~eV) lines are also
generally visible when Z And is inactive,
suggesting $T_{hot} \gtrsim 10^5$ K in quiescence.
During the 1997 outburst, Raman $\lambda$6830 strengthened considerably,
confirming the increase in $T_{hot}$ to 180,000~K suggested by
$F_{4686} /F_{H\beta}$.
During the rise to optical maximum in 2000, 
[\ion{Fe}{7}] $\lambda$6087 and Raman $\lambda$6830 weakened
and then disappeared, and in {\it FUSE} spectra, O VI $\lambda
\lambda$ 1032,1038 was absent, all supporting the decline in $T_{hot}$
to $\sim$ 80,000--90,000~K.
These lines reappeared during the optical decline, confirming the
rise in effective temperature
determined from $F_{4686} /F_{H\beta}$.
In $FUSE$ spectra F4 -- F6, taken midway through the optical decline
in 2001, \ion{O}{6} $\lambda \lambda$ 1032,1038 are strong, but
\ion{Ne}{6} $\lambda \lambda$ 1006,1010 ($\chi = 126$~eV) are weak or
absent.
For this time period, we adopt $120,000$~K, slightly below our
$F_{4686} /F_{H\beta}$ estimates.  In $FUSE$ spectra F7 and F9, in 2002
July and 2003 August,
\ion{Ne}{6} $\lambda \lambda$ 1006,1010 are present, confirming the
increase in $T_{hot}$ back to the pre-outburst value, in accord with
the simultaneous return to the optical quiescent state.  During the
brief optical rebrightening in late 2002, the fading of \ion{Ne}{6}
$\lambda
\lambda$ 1006,1010 on $FUSE$ spectrum F8 indicates that $T_{hot}$
temporarily decreased, and we again adopt a value of $120,000$~K for
this time period.  Overall, our line-ratio-method and
ionization-state-based estimates of $T_{hot}$ 
appear to be consistent to within 10-20\%.
The 
hot-component effective temperature, $T_{hot}$, is plotted
as a function of time in the
top panel of Fig.~\ref{fig:tempev}.

We found that
$T_{hot} = 150,000 \pm 7,000$~K during quiescence.  Our value of the quiescent WD
effective temperature is somewhat higher than the quiescent-state
Zanstra temperature of 120,000~K (on average) found by
\cite{mur91} between 1979 and 1987, and of roughly $10^5$ K found by
\cite{fc88} between 1978 and 1982.  
However, the effective temperatures from \cite{mur91} increase during
1979--1981 (108,000~K on average) and 1984--1987 (126,000~K on
average), and \cite{birrielea98} find $T=111,000 \pm 4,000$ K from a
FUV observation in 1995 March.
Our higher values could thus 
reflect long-term changes of a similar magnitude in $T_{hot}$.
\cite{tomov03} find $T_{hot} = 150,000$ K, as we do, in 1999 September.

In summary, the effective-temperature history of the WD is markedly
different for the 1997 and 2000--2002 outbursts.  During the 1997
outburst, $T_{hot}$ increased by about 20\%, to more than 180,000 K.
It returned to the
quiescent-state value by midway through the optical
decline. $T_{hot}$ showed a more complex evolution during the
2000--2002 event.  
Photometry from \cite{sko2003} indicates that very early in the
2000--2002 event (within the first few days), the $U$-band flux jumped
by 2 mag and the $U-B$ color
decreased
\cite[i.e., became bluer; see][for more complete coverage of this blue
spike]{sko2003}, suggesting that $T_{hot}$ initially increased.
Within two weeks of the start of the outburst, however, as the optical
flux continued its rise to maximum, the $T_{hot}$ evolution changed
direction.  Over the next two months, $T_{hot}$ decreased to $<$
90,000 K.  By the time of optical maximum, $T_{hot}$ was beginning its
recovery.  It returned to the quiescent-state value before the
completion of the optical outburst.  Finally, $T_{hot}$ again dropped
briefly during the small optical rebrightening in late 2002.  The
dominant behavior of $T_{hot}$ dropping as the optical flux rises to
maximum, as in the 2000--2002 event, is typical of historical
outbursts.

\subsubsection{Luminosity} \label{sec:lum}

To determine the bolometric luminosity of the WD throughout the
2000--2002 outburst, we estimated the WD radius at the time of each
$FUSE$ observation by scaling the WD photosphere models of
\cite{barman00} to the extinction-corrected $FUSE$ fluxes, using
the effective temperatures described above, a distance of
1.2~kpc,
and a mass of 0.65 
$\msun$ \citep{schsch97}.
Values of $E(B-V)$ in the literature include 0.3 mag
\citep{alt81}, 0.27--0.29 mag \citep{kw84}, 0.3 mag \citep{mk96}, and
0.35 mag
\citep{viotti82}.  We take $E(B-V)=0.27$ mag, the lowest of these values.
This choice is the most conservative for our purposes, as larger
values will produce even greater WD bolometric luminosities.
Furthermore, this low value of $E(B-V)$ produces a dereddened
flux ratio ($F_{\lambda 1059}/F_{\lambda1103}$) that is more
consistent with the theoretical expectation, and also more consistent
with the interstellar reddening from dust emission maps
\citep{schlegelea98,bh82}. 
With the WD photospheric radii and effective temperatures determined
previously, we were then able to obtain the luminosities using the
standard relation $L_{hot}=4\pi R_{hot}^2 \sigma T_{hot}^4$.  The $FUSE$
fluxes provide a good approximation to the flux directly from the WD
because the contribution from both the red giant and the nebula are
negligible in the FUV \citep[for a complete spectral energy distribution
and model, see][]{fc88}.

The derived hot-component luminosities, which we list in
Table~\ref{tab:fuvlums}, are close to $10^4\, \lsun$ from the end of
2000 to late 2001.  Although these results are quite sensitive to the
reddening used to correct the FUV fluxes for extinction, our choice
for $E(B-V)$ on the low end of the range of published estimates means
that the true luminosities could have been even higher.  In addition,
we find reasonable agreement between our estimate of $L_{hot}
\approx 1500$--$2000\, \lsun$ in quiescence ($FUSE$ observations F7
and F10) and the quiescent-state $L_{hot}$ estimates of approximately
620--$1600\, \lsun$ found by \cite{mur91}, who used a
modified Zanstra method,
which is much less sensitive to reddening.  For comparison with our
outburst luminosities, the Eddington luminosity for a $0.65\, \msun$
white dwarf is $L_{Edd} = 3 \times 10^4\, \lsun\, (M/0.65\, \msun).$

\subsection{Disk Instability in 1997} \label{sec:diin97}

We interpret the 1997 outburst in terms of a disk instability
that produces a sudden increase in the accretion rate through the disk
surrounding the white dwarf. Considerable evidence supports this
interpretation:
1) the amplitude and shape of the optical light curve; 
2) the behavior of the short-period photometric oscillation
due to magnetic accretion onto a rotating white dwarf; and
3) the evolution of $T_{hot}$ in 1997.

Given the small size and short duration of the 1997 outburst, it
was unlikely to have been caused by a recurrent-nova-like
thermonuclear runaway or a weak shell flash.  Thus, the main alternative
to the disk-instability interpretation is an expansion of the
white-dwarf photosphere at roughly constant bolometric luminosity
\citep{tutuyun}.
In symbiotic stars, \cite{kw84} showed how the spectroscopic evolution
of this alternative differs from the evolution expected from a disk
instability. Here, we consider each of the pieces of evidence listed
above and use this information to reject the expanding-photosphere
picture in favor of a disk instability.

The triangular shape of the light curve during the 1997 outburst
resembles the outbursts of many dwarf novae
\citep[see, e.g.,][]{warnerbook}. The rapid rise time of $\lesssim$ 20 days
is similar to the typical rise times of $\sim$ 2--10 days
for dwarf novae, especially given that longer dwarf-nova rise
times are typically seen in CVs with longer orbital periods \citep[and
therefore larger disks;][]{warnerbook}.  The slower, nearly linear
decline is also reminiscent of dwarf novae, where a cooling wave
propagates inward through the disk at the local sound speed. In dwarf
novae, the cooling wave takes, for example, 7 -- 10 days to travel roughly
0.7 R$_{\odot}$. The longer 300 day decline in Z And is reasonable for
a disk with an outer radius of 20 -- 40 R$_{\odot}$, or roughly 3 -- 7\%
of the binary separation (10 -- 20\% of the hot-component Roche-lobe
radius if we take $M_{WD} = 0.65\, \msun$ and a red-giant mass of
$1.3\, \msun$).  Moreover, the faster rise than decline is suggestive
of an outburst in which the heating wave is initiated at large radii
and moves inward through the disk \citep{cwp86}.  Besides the shape, the
amplitude
of the 1997 outburst is close to the $\sim$1 mag maximum amplitude
expected for a disk instability in a symbiotic star
\citep[the amplitude is low because of the additional light from the
red giant and 
nebula;][]{kbook,duschl86b}.

During the small 1997 outburst, \cite{sb99} detected a
28-min photometric oscillation 
when the system was near optical maximum and then throughout the
decline back to optical quiescence.  
They interpreted this oscillation as a hot-spot on the
surface of a white dwarf with a rotational period of 28 min. The
properties of the light curve are consistent with a hot spot produced
by accretion channeled from the disk to the white dwarf by a magnetic
field with a surface strength of at least $3 \times 10^4$ G (or $6
\times 10^6$ G if the WD is in spin equilibrium). The behavior of the
oscillation throughout the outburst provides a clear discriminant
between a disk instability and an expanding photosphere. In the
expanding-photosphere model, the optical emission from accretion
remains constant while optical emission from the WD photosphere
increases.  Therefore, the ratio of optical flux from accretion to
that from the WD photosphere drops by at least the same factor as the
overall increase in optical brightness.
Thus, the expanding-photosphere model predicts that the pulse
fraction of the oscillation will decrease during an outburst.
In contrast, because the luminosity
of an accretion hot-spot is likely to track the overall accretion
luminosity, the pulse fraction of the oscillation should remain
roughly constant during a disk instability.

Our data clearly favor a disk instability as the
source of the 1997 outburst.
The pulse fraction of $\sim$ 0.6\% near the peak of the outburst is
comparable to the pulse fraction of $\sim$ 0.7\% during quiescence
\citep[after subtracting out the approximate contribution of the red giant
to the $B$-band flux, which we take to be 75\% in quiescence; see][]{fc88}.
In contrast, we would expect at least a factor of three reduction
in the pulse fraction for a 1 mag rise in brightness due to an
expanding photosphere.
The data clearly rule out such a large change in the pulse
fraction.

The derived change in $T_{hot}$ also favors a disk instability over an
expanding photosphere. In the Rayleigh-Jeans tail of the WD blackbody
distribution, the optical intensity is proportional to $T_{hot}$.  The
optical flux is therefore proportional to $R_{hot}^2 T_{hot}$. If the
bolometric luminosity is constant, as is the case in the
expanding-photosphere model, then $R_{hot}^2
\propto T_{hot}^{-4}$, and the optical flux is proportional to
$T_{hot}^{-3}$.  For a 1 mag optical eruption, the
expanding-photosphere model thus predicts a factor $\sim$ 1.3
decline in $T_{hot}$ \citep[see also][]{bathshaviv76}.
Our analysis of the optical emission lines
indicates a factor of 1.2 
increase in
$T_{hot}$ (Fig.~\ref{fig:tempev}).  The disk-instability model is
clearly more consistent with the data.

\subsection{Disk Instability Plus Enhanced
Shell Burning in 2000--2002} \label{sec:nucpower}

The hot-component effective-temperature evolution 
was quite different during in the 2000--2002 outburst.
These two eruptions were therefore
examples of distinct types of events.
Unlike the 1997 event, the properties of the 2000--2002 outburst are
not consistent with a simple disk-instability model.  Evidence against
the simple disk-instability model for this eruption includes the following:
1) the rise to optical maximum proceeded in three distinct stages;
2) the optical brightness increased by well beyond the 1 mag
that can be produced by a disk instability in a symbiotic
\citep[][]{kbook};
3) $T_{hot}$ evolved in a rather complex way;
4) the optical oscillation at the WD spin period due to magnetic
accretion was  
absent near the peak of the outburst; and
5) the amount of material
needed to power this event by accretion alone is much higher than
could reasonably have been accreted during the outburst.

\begin{figure*}[t]
\begin{center}
\epsfig{file=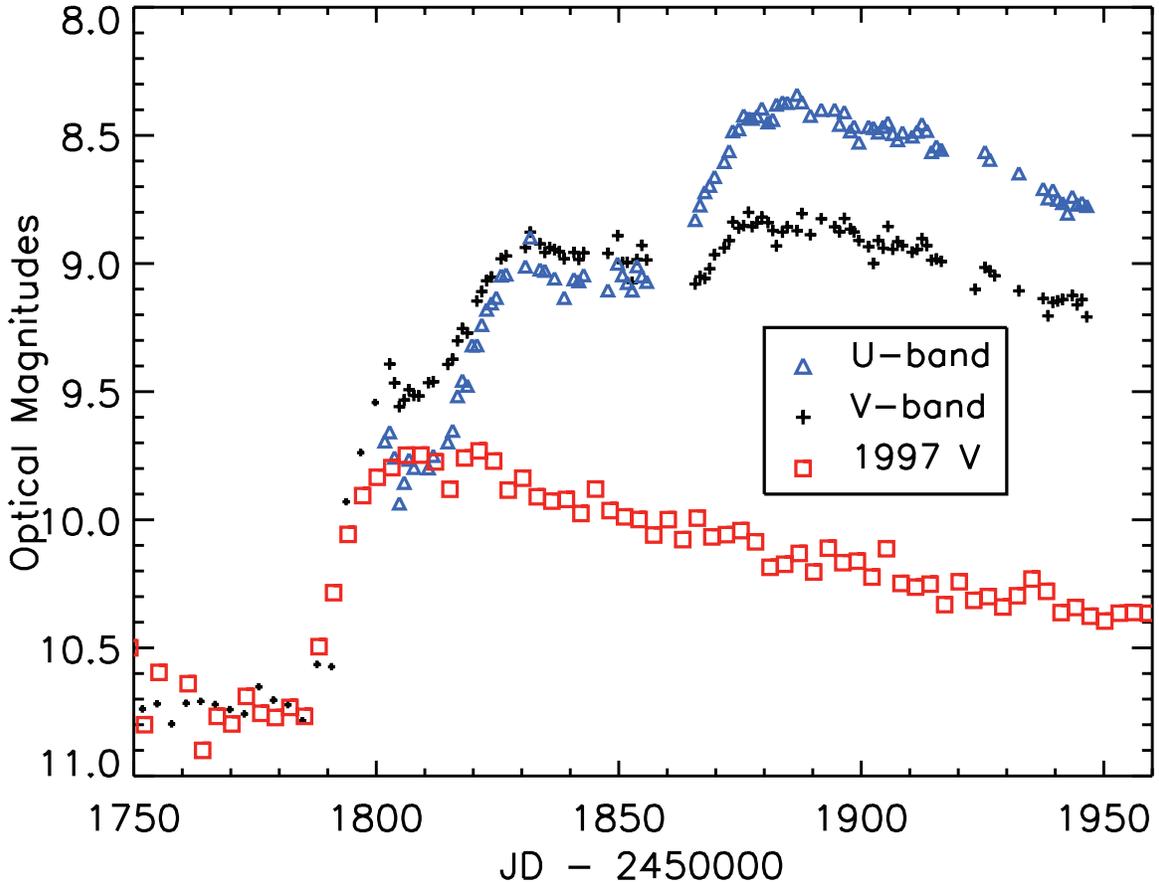,width=6in}
\end{center}
\caption{$U$-band (blue triangles) and $V$-band (black dots and crosses)
light curves from the early part of the 2000--2002 outburst of Z And,
with the 1997 $V$-band light curve (red squares) shifted in time and
overplotted.  The optical rise initially follows the same course for
the two events.  In 2000, however, the first-stage rise proceeds
beyond the maximum level reached in the 1997 event.
The outburst evolution subsequently takes a different path.
The blue triangles and black crosses are data from KAIT, and the
black dots and red squares are data from the AAVSO. \label{fig:oplot}}
\end{figure*}

The start of the 2000 eruption closely resembled the evolution of the
1997 outburst.
Fig.~\ref{fig:oplot}, in which the 1997 and 2000 light curves are
overlaid, shows the similarity between the 
first stage of the 2000--2002 event and the
rise to maximum in 1997.
Very early in the 2000--2002 event, $T_{hot}$ appears
to have increased, as it did in 1997 (see \S\ref{sec:tempev}).
We do not have the spectral data needed to 
determine $T_{hot}$ during the first week or so of the 2000--2002
outburst, or the fast photometry to show that the pulse fraction of
the oscillation remained constant during the early rise in
2000. However, the $U-B$ color evolution during the first week of the
outburst \citep{sko2003} strongly suggests that $T_{hot}$ increased
during the early rise.  Furthermore, since the $U$ brightness tends to
be dominated by reprocessed ionizing radiation from the hot component,
the early jump in $U$-band flux \citep{sko2003} indicates that the
2000--2002 outburst was initiated by a process that caused the
luminosity of the hot component to increase rapidly.
We therefore conclude that the 2000--2002 outburst began with a disk
instability, as in 1997.  

If the 2000--2002 outburst was entirely accretion powered, the
required accretion rate to produce such a high luminosity would be
\begin{eqnarray}
\mdot & = & \frac{L R_{hot}}{G M_{WD}} \gtrsim 10^{-5} 
\frac{L}{10^4\, \lsun} \frac{R_{hot}}{0.1\, \rsun} \times  \nonumber \\
 & & \hspace{2cm} \left(\frac{M_{WD}}{0.65\, \msun} \right)^{-1}\, \msyr,
\end{eqnarray}
where $L$ is the bolometric luminosity, $R_{hot}$ is the WD radius,
$M_{WD}$ is the WD mass, 
and 
$R_{hot} = 0.1\, \rsun$ as indicated by \cite{mur91} and our scaling of
the FUV fluxes to photospheric models (see Table~\ref{tab:fuvlums}).
It is difficult to imagine how such a high accretion rate could be
sustained in Z And for a full year.  

If the outburst was nuclear powered, on the other hand, the burning of
only a few times $10^{-7}\,
\msun$ of fuel can produce $L_{hot} \sim 10^4\, \lsun$ for one year.
Moreover, since $L_{hot}$ was higher by an order of magnitude in
outburst compared to quiescence, and we argued in \S~\ref{sec:intross}
that the quiescent hot-component luminosity already indicates the
presence of nuclear shell burning, it follows that the rate of nuclear
burning must have increased during outburst.
\cite{kilpio05} also suggest that
this outburst involved enhanced nuclear burning (but as a result of
the collapse of the disk when the wind speed dropped below a critical
level).  If we include the kinetic energy of ejected mass in the total
energy produced by the outburst
\citep{brock04,tomov03}, thermonuclear 
involvement is even more strongly indicated.

\subsection{Shell Ejection and Dissipation}

Our observations suggest that during the second stage of the
2000--2002 outburst, a shell of material (or possibly an optically
thick wind) was blown from the surface of the WD.  During the third
stage,
this ejected shell became optically thin.  During the second stage,
$T_{hot}$
dropped (see Fig.~\ref{fig:tempev}) and the $U-B$ color
reddened, first sharply, and then more gently (see
Fig.~\ref{fig:kaitfig}).
The $FUSE$ spectrum taken
near the end of the second stage was
dominated by blueshifted absorption from optically thick gas flowing
outward (see \S\ref{sec:obs_fuv}).  Instead of the
high-ionization-state emission lines typically associated with
symbiotic-star nebulae, this first $FUSE$ observation shows a large
amount of cool gas (e.g., singly and doubly ionized species of C, Fe, and
Si).  The hot WD surface thus appears to have been hidden by
an optically thick shell of material.
During the third stage of the rise to optical maximum,
$T_{hot}$ slowly began to increase again and the $U-B$ color became
more blue, the final increase in optical flux became dominated by
$U$-band light,
the FUV spectra began to move from absorption to emission, and the radio flux
rose.
After the third stage, the X-ray spectrum showed evidence for the
presence of significant quantities of shock-heated gas.

The uncharacteristically low radio flux density measured during the
second stage of the outburst
(see \S\ref{sec:obs_radio}) is consistent with the ejection of an
optically thick shell from the WD.
The radio flux is primarily due to 
bremsstrahlung emission from the nebula.  A low value indicates that
the size of the emitting region is unusually small, in this case
presumably because the flux of ionizing photons decreased when the
shell first expanded and 
$T_{hot}$ dropped.  At the time of the first radio observation (R1),
we estimate $T_{hot} \lesssim 89,000\, {\rm K}$.  We
thus expect the flux of
ionizing photons from the hot-component surface to drop more
than 80\% from the pre-outburst value 
to $n_{ioniz}(R1) = 0.16\; n_{ioniz,q}$, where
$n_{ioniz}(R1)$ is the flux of ionizing photons from the
hot-component photosphere 
at the time of R1, and
$n_{ioniz,q}$ is the flux of ionizing photons in quiescence, when
$T_{hot} \approx 150,000$~K.

Given a typical quiescent 5 GHz radio flux density at the orbital
phase corresponding to R1 of $S_{5\,GHz, q} = 0.9 \pm 0.3$~mJy
\citep{kennyphd}, and a measured 5 GHz radio flux density
of $S_{5\,GHz}(R1) = 0.42
\pm 0.06$~mJy, 
we can infer the approximate radius of the hot-component photosphere
or ejected shell:
\begin{eqnarray}
R_{hot}(R1) & \approx & R_{hot,q} \left( \frac{S_{5\,GHz}(R1)}{S_{5\,GHz,q}}
\right)^{1/2} \left( \frac{n_{ioniz,q}}{n_{ioniz}(R1)} \right)^{1/2}
\nonumber \\
& \approx & 0.17 \pm 0.06\, \rsun,
\end{eqnarray}
where we have taken the quiescent-state hot-component radius to be
$R_{hot,q} \approx 0.1\, \rsun$ \citep[see][]{mur91}.
We have assumed
that changes in the size of the ionized portion of the nebula occur in
the outer regions that are closer to optically thin than optically
thick.  Thus, the change in radio flux density is approximately equal
to the change in total flux of ionizing photons.  Since the geometry
and ionization structure of the nebula are poorly known, the resulting
uncertainty in our $R_{hot}$ estimates from the radio data is probably
even larger than the quoted errors.
We expect the change in flux of ionizing photons to impact the radio
emission almost immediately, since the recombination time in a nebula
with electron density on the order of $10^{9}$--$10^{10}\, {\rm
cm}^{-3}$ is minutes to hours \citep{fc95}.

A similar analysis of the radio flux at the time of our second 
radio observation (R2) gives $R_{hot}(R2) = 0.23 \pm 0.09\, \rsun$ at JD
2451865, roughly consistent with
$R_{hot}$ derived from a $FUSE$ observation taken at the same time (see
\S\ref{sec:nucpower} below).
From analysis of additional $FUSE$ observations taken between the
second and third radio observations, we find that $R_{hot}$ reached a
maximum size of greater than $0.35\, \rsun$.  The third radio
observation indicates that the hot-component photosphere had
contracted, to $R_{hot}(R3) = 0.27
\pm 0.1\, \rsun$, by JD 2451921.
Thus, the expansion and dissipation
of the photosphere or optically thick shell is evident in the radio as 
well as FUV data.

Our first radio data point was obtained during the optical plateau
after the second optical rise, over one month into the outburst.
Therefore, the only information about the evolution of
$R_{hot}$ very early in the outburst is that indirectly inferred
from the decline in $T_{hot}$ between optical spectroscopic
observations O1 and O3.  A
radio light curve (plus simultaneous optical spectroscopic monitoring)
starting earlier in a classical symbiotic-star outburst would provide
more complete information about the evolution of $R_{hot}$ during the
beginning of the shell-ejection phase of this type of event.  Since
the FUV fluxes can provide an independent measure of $R_{hot}$, a
simultaneous FUV light curve early in an outburst would provide a test
of the shell-ejection hypothesis.

The non-detection of the magnetic
hot spots on the surface of the WD
at the beginning of the third stage of the outburst  
is also consistent with the ejection of an
optically thick shell from the WD.  Because the WD in Z And is
magnetic, the accretion flow close to the WD is channeled along
magnetic field lines.  Hot spots at the base of these columns produce
an optical oscillation at the WD spin period \citep{sb99}.  However,
on JD 2451872 (just before the optical maximum), the oscillation
amplitude upper limit was well below the amplitude measured during the
1997 outburst, suggesting that
the white-dwarf surface was either hidden or the magnetic accretion
flow disrupted during the second stage of the rise to optical maximum
in late 2000.

Our X-ray observations are also consistent with the shell-ejection
hypothesis.  The hard ($>$2 keV) tail seen in the second X-ray
observation (X2) could be the result of shock-heated gas, which in
symbiotic stars has been suggested to be due to colliding winds
\citep[e.g.,][]{murset97}.
A new burst of material blown from the WD would tend to enhance
such emission.  Furthermore, as discussed in \S\ref{sec:obs_xray}, at
least one of the absorption edges we detected can be ascribed to an
ionization state of N that is most likely produced in shock-heated
gas.  The difference between the 1997 and 2000 peak X-ray fluxes
provides additional support for shell ejection during the 2000--2002
event.  In 1997, no shell was ejected, and so the accreting WD was
clearly visible.  Accordingly, the X-ray flux reached a significantly
higher value in 1997 compared to 2000, when much of the X-ray flux
from accretion was probably hidden.

The physical changes that comprise the third stage of the rise are,
however, reflected most clearly in the
FUV spectra.
During this period, the FUV emission begins to leak through the
ejected shell, producing the high FUV fluxes seen in 2000 November and
December and the $U$-dominated final optical rise (via nebular
reprocessing of FUV emission into the optical).
The evolution of the FUV spectrum from an optically thick, absorption-line
spectrum with blueshifted P-Cygni absorption to optically thin
emission-line spectra provides the most direct evidence for shell
ejection and dissipation.
Moreover, the evolution of the blueshifted P-Cygni absorption reveals
that the average velocity of the absorbing material along the line of sight
decreased with time.  
Finally, the maximum velocities from the first $FUSE$
observation are comparable to the 400 km s$^{-1}$ inferred for the
speed of the radio jet that was produced in conjunction with the
2000--2002 outburst \citep{brock04}.

\subsection{Optical Rebrightening in 2002} \label{sec:rebright}

In mid-2002, the decline to optical quiescence was 
interrupted by a $\sim$1 mag rebrightening that lasted for about five
months.  
The $U-B$ color during this rebrightening was initially quite red
(comparable to 
the second stage of the outburst; see bottom panel of
Fig.~\ref{fig:kaitfig}).  As the rebrightening event faded, the $U-B$
color became slightly more blue, although never as blue as during the
entire previous year-long optical decline.  Both the $FUSE$ and VLA
fluxes were unusually low 
during the optical rebrightening.

Our data indicate that the optical rise in 2002 could have
resulted from a decrease in $T_{hot}$ due to a slight expansion of the
white-dwarf photosphere. 
\cite{sko2003} proposes that
the rebrightening was due to the emergence of the hot component from
eclipse.
The unusually low values of the FUV and radio fluxes near the maximum
of the optical rebrightening, and the orbital inclination of $47 \pm
12^\circ$ from polarimetry \citep{schsch97}, both present problems for
the eclipse interpretation.
Moreover, the fact that the optical flux is higher after the dip in
the light curve despite the system being at an orbital phase where it
is normally low,
and that the minimum of the possible eclipse occurs before orbital
phase zero (see 
Fig.~\ref{fig:tempev}), also
tend to favor a non-eclipse explanation.

\cite{webbink76} suggested that a similar rebrightening in T CrB was
due to the emptying of the accretion disk after a pulse of material
ejected from the red giant formed a ring and then disk around the
accreting star.  
For Z And, however, the disappearance of the \ion{Ne}{6} lines in
$FUSE$ spectrum F8,
the reddening of the $U-B$ color, and the decrease in the radio flux
density all indicate that $T_{hot}$ decreased (from around 150,000~K
to approximately 120,000~K) when the optical flux increased.
To produce this decrease in $T_{hot}$, we infer that the WD radius
probably expanded slightly, despite the fact that our FUV data do not
allow a reliable estimate of $R_{hot}$ at this time (the FUV flux
ratio $F_{1059}/F_{1103}$ indicates that our FUV continuum fluxes in 
observations F7--F9 must
have been contaminated by emission and/or absorption, so a WD
photosphere model could not be fit to derive WD radii).
\cite{tomovea04}
report a similar conclusion regarding the optical activity in 
2002.

\section{Discussion} \label{sec:implications}

In the 2000--2002 eruption of Z And, we see evidence for phenomena
that usually occur in two distinct types of CV outbursts (dwarf novae
and classical novae).  As discussed in \S\ref{sec:nucpower}, there are
multiple similarities between the first stage of the 2000 outburst and
the 1997 event, which was due to an accretion-disk instability (see
\S\ref{sec:diin97}).  The 2000--2002 outburst thus appears to have
started as an accretion-disk instability.  After a few weeks to
months, however, the outburst energetics became dominated by an
increase in nuclear shell burning on the surface of the WD.  In
addition, during the rise to optical maximum, an optically thick shell
of material was ejected, causing $T_{hot}$ to drop and P-Cygni
profiles to appear in the $FUSE$ spectra.
We suggest that the 2000--2002 outburst of Z And was triggered by the
influx of hydrogen-rich material from a dwarf-nova-like disk
instability, and then powered by the subsequent increase in nuclear
shell burning on the WD, in a milder version of the phenomenon that
powers classical novae.
The outburst in Z And therefore combines elements of dwarf novae and
classical novae.  We refer to this new type of eruption as a {\it
combination nova}.  Since the latter stages of typical
classical symbiotic outbursts are similar to the 2000--2002 outburst
of Z And, and the the recurrence times for classical symbiotic
outbursts are shorter than theoretical calculations can obtain for
standard shell flashes \citep{sionea79}, it is possible that many
classical symbiotic outbursts are combination novae.

The initial correspondence between the 2000 and 1997 outbursts
suggests that the physical trigger mechanism was the same for both
events.
A disk-instability trigger mechanism can explain the recurrence times
(from a few years to more than a decade) of the outbursts of Z And and
other classical symbiotic stars.  Furthermore, since symbiotics are
wide binaries, their accretion disks are probably large.  For
typical accretion rates of $\sim 10^{-9}$ to $\sim 10^{-8}
\msyr$, the disk temperature ranges from $\sim 40,000$ K close to the
white dwarf surface to $\lesssim 1$-- $2 \times 10^3$ K at $\sim 3\,
\rsun$.  These conditions encompass the range needed for an unstable
disk
\citep[e.g.,][and references therein]{cannizzo93,osaki96}.
Based on a correlation between the presence of classical symbiotic
outbursts and either observational 
indicators that the red giant is close to filling its Roche lobe (and
therefore that a disk is more likely to be present around the WD) or
photometric oscillations similar to the superhumps in the SU UMa class 
of CVs,
\cite{mska02} suggest that the presence of an accretion disk drives
the eruptions in the symbiotic binaries Z And, YY Her, CI Cyg, BF Cyg,
and AX Per.  Our extensive observations of Z And show that this
assertion is true for this object, and explain how the presence of an
accretion disk can produce classical symbiotic outbursts.

For the disk-instability trigger mechanism to produce an observable
increase in the rate of nuclear shell burning on the WD,
the amount of material added, $\Delta M_{DI}$, must be
a significant fraction of the mass of the WD envelope, 
$M_{env}$.
Extrapolating slightly from the curves in Fig.~8 of \cite{tb04}, we
expect the envelope mass on a $0.65\, \msun$ WD accreting at a rate of
a few times $10^{-8}\, \msyr$ to be around $2
\times 10^{-5}\, \msun$.  
For the luminous WD in Z And, the envelope mass should be even smaller
since less material is required to initiate a thermonuclear runaway or
weak shell flash on a luminous WD 
\citep[e.g.,][]{sionstar86}. 
Observationally, Fig.~\ref{fig:oplot} shows that the first stage of
the 2000 event rose higher than the maximum reached in 1997,
apparently crossing some threshold in $\Delta M_{DI}+M_{env}$ needed
to produce a thermonuclear response.  For the 1997 event, which we
assume was entirely accretion powered, if we take a bolometric
correction $BC = -2$ mag,
we find $\Delta M_{DI}$ is approximately equal to a few times
$10^{-7}\, \msun$.  The amount of material accreted in 2000
was probably similar or somewhat larger.  Thus $\Delta
M_{DI}$/$M_{env} \approx 0.01$--0.1 (for $\Delta M_{DI} \approx
10^{-7}$--$10^{-6}\, \msun$)
appears to be sufficient to produce enough of an increase in pressure,
and hence temperature, to trigger a nuclear response in the burning
shell of Z And.
The relatively small amount of accreted material that appears to be
able to trigger a shell flash, and the correspondingly short amount of
time between flashes, cannot be explained by current models of nuclear
burning on the surface of a WD.  Episodic accretion through an
unstable disk could be the additional factor that needs to be taken
into account.  If we take an average accretion rate into the WD disk
of $5
\times 10^{-8}\, \msun$
\citep[see][who finds typical mass transfer rates
in symbiotic stars of $10^{-7}$ to $10^{-8}\, \msyr$]{iijima02},
then it would be possible to both fuel the quasi-steady burning and
accumulate enough material in the disk to trigger a combination nova
roughly every ten years.

After the addition of fresh fuel, we expect the nuclear burning rate
to increase on the thermal time scale at the base of the envelope, 
\begin{equation}
t_{th} = \frac{C_P T \Delta M}{L_{hot}} = \frac{C_P T}{E_{nuc}}\frac{\Delta
M}{\mdot},
\end{equation}
where $C_P$ is the specific heat at constant pressure, $T$ is the
temperature at the base of the WD envelope, $\Delta M$ is the mass of
the envelope, and $E_{nuc} = 5 \times 10^{18}$ erg g$^{-1}$ 
is the amount of energy released by burning one gram of hydrogen-rich
material.  For steady CNO burning, the temperature at the base of the
envelope is $\sim 3 \times 10^7$ K.  However, since the CNO
burning rate, $\epsilon_{nuc}$, is a very sensitive function of $T$
($\epsilon_{nuc} \propto T^{20}$), 
and this burning rate increased by a factor of ten during the 2000--2002
outburst of Z And, the observed burning enhancement can be produced by
about a 10\% change in the temperature at the base of the envelope, or 
a change of less than $0.4
\times 10^7$ K.  Thus, the
thermal time scale associated with the observed degree of change in
nuclear burning in Z And is
\begin{eqnarray} \label{eqn:tth}
t_{th} & \approx & 1\, {\rm month} \left(\frac{T}{4 \times 10^6\, {\rm
K}} \right) 
\left( \frac{\Delta M}{2 \times 10^{-5}\, \msun} \right) \times
\nonumber \\ 
 & & \hspace{3cm} \left(
\frac{\mdot}{5 \times  
10^{-8} \msyr} \right)^{-1}.
\end{eqnarray}
Here, we have taken an envelope mass of $2 \times
10^{-5}\, \msun$, 
and an average accretion rate of $5 \times 10^{-8}\, \msyr$.  

The second stage of the 2000--2002 outburst occurred about 2 weeks
after the beginning of the outburst.  This response time is quite close
to the expected nuclear response time given in Eq.~\ref{eqn:tth},
which could in fact be even shorter if the total amount of fresh fuel
is a significant fraction of the WD envelope and is dumped onto the
WD quickly (i.e., within a few days to a week), so that the compression
becomes adiabatic.  
To test this disk-instability-trigger idea
for future symbiotic-star outbursts, it will be important to obtain
optical spectra during the first few days of the initial rise.

Both symbiotic stars and supersoft X-ray sources probably
contain accreting WDs with nuclear shell burning
\citep{vdh92,mur91,sbh01}.
However, based upon the lower hot-component luminosity
estimates for symbiotics, the burning may occur at a slower rate in
symbiotics (and in a few, it is completely absent).  The
cause of optical and X-ray brightness changes appears to be another
important difference between these two classes of objects, both of
which are considered to be possible progenitors of Type Ia supernovae.

A promising model for the long-term variability of at least some
supersoft X-ray sources is the simple photospheric-expansion model (in
which the WD bolometric luminosity remains constant), since X-ray
fluxes have been seen to drop when the optical fluxes rise in several
supersoft sources \citep[e.g.,][]{southwell96}.  We do not observe this inverse
relationship between optical and X-ray flux during either the 1997 or
the 2000--2002 outbursts of Z And.
As discussed in \S\ref{sec:templumev} and
\S\ref{sec:nucpower}, the rise in $T_{hot}$ 
in 1997 and the increase in $L_{hot}$ by an order of magnitude during the
2000--2002 outburst are also at odds with the simple photospheric-expansion
model.  Finally, given the low (compared to supersoft X-ray sources)
hot-component effective temperature for Z And, blackbody emission from 
the WD surface probably plays very little role in producing the X-ray
emission from this source.   
Instead, the X-ray observations of Z And are consistent with the
combination nova picture.

\section{Summary} \label{sec:sum}

We have shown that the 2000--2002 outburst of Z And was primarily
powered by an increase in nuclear burning on the WD surface.  The
smaller, 1997 event, on the other hand, is consistent with 
having been a
disk instability like those of dwarf novae.
Thus more than one type of classical symbiotic outburst can occur
within the same symbiotic system.  The key points that allow us to
gain new insight into classical symbiotic outbursts from the
2000--2002 eruption of Z And are as follows: 1) high-precision $UBV$
light curves that showed the outburst to be a multistage event; 2) the
close similarity between the first stage of the multi-stage outburst
and the 1997 outburst, which we identify as a dwarf-nova-like disk
instability; and 3) the fact that the hot-component luminosity during
much of the eruption, as determined from FUV fluxes, was too high for
the outburst to have been solely accretion-driven.

In the 2000--2002 event, the bolometric luminosity increased
during the outburst to $\sim 10^4 \lsun$, indicating that
the outburst could not have been
due to either a simple expansion of the white-dwarf photosphere (e.g.,
in response to a small change in $\dot{M}$ to above the steady-burning
upper limit) or to a burst of accretion.  Instead, the overall
energetics support 
thermonuclear shell burning as the main source of power. However,
since both the short recurrence time for large outbursts in Z And and
the structure of the 2000--2002 outburst light curve
suggest that the increase in nuclear burning 
was not a standard recurrent nova,
we suggest that the enhanced nuclear burning was triggered by an
accretion event.  We favor a dwarf-nova-type disk instability as the
trigger.  Thus, we propose a new type of outburst that
combines the physics of dwarf novae and classical novae.  We call this
type of event a {\it combination nova}.

\acknowledgments

We are grateful to L. Bildsten for useful discussions, to C. Crowley
for assistance with the $FUSE$ data analysis, and to R. Hynes for
providing WHT spectra.  The referee, S. Starrfield, also provided many
suggestions that improved the paper.  We acknowledge the
variable star observations from the AAVSO International Database
contributed by observers worldwide and used in this research.  IRAF is
distributed by the National Optical Astronomy Observatory, which is
operated by the Association of Universities for Research in Astronomy,
Inc. under contract to the National Science Foundation.  J.L.S. is
supported by an NSF Astronomy and Astrophysics Postdoctoral Fellowship
under award AST 03-02055.  The work of A.V.F.'s group at
U. C. Berkeley is supported by National Science Foundation grant
AST-0307894.  KAIT was made possible by generous donations from Sun
Microsystems, Inc., the Hewlett-Packard Company, AutoScope
Corporation, Lick Observatory, the National Science Foundation, the
University of California, and the Sylvia \& Jim Katzman Foundation.
The National Radio Astronomy Observatory is a facility of the National
Science Foundation operated under cooperative agreement by Associated
Universities, Inc.

\clearpage


\begin{deluxetable}{cccc}
\tablewidth{0pt}
\tablecaption{Optical Photometry \label{tab:optphot}} 
\tablehead{
\colhead{Date} & \colhead{$U$} & \colhead{$B$} & \colhead{$V$} \\
\colhead{(JD)} & \colhead{(mag)} & \colhead{(mag)} & \colhead{(mag)}}
\startdata
2451801.75 & 9.69   & \nodata & \nodata \\
2451802.75 & 9.66   & 10.41 &  9.39 \\
2451803.75 & 9.76   & 10.39 &  9.47 \\
2451804.75 & 9.94   & 10.61 &  9.56\\
2451805.75 & 9.86   & 10.59 &  9.53\\
2451806.75 & 9.76   & 10.54 &  9.49\\
2451807.75 & 9.80   & 10.57 &  9.52\\
2451808.75 & \nodata& 10.59 &  9.52\\
2451810.75 & 9.80   & 10.44 &  9.47\\
2451811.75 & 9.75   & 10.37 &  9.46\\
$\vdots$ & $\vdots$ & $\vdots$ & $\vdots$ \\
\enddata
\tablecomments{The complete version of this table is in the
electronic edition of the Journal, and includes photometry from 276
nights.  The printed edition contains only a sample.  The relative
photometry is accurate to approximately 1\%.  Absolute photometry is
accurate to approximately 10\%.}
\end{deluxetable}

\begin{deluxetable}{cccc}
\tablewidth{0pt}
\tablecaption{Selected Optical Spectroscopic Observations
\label{tab:optspec}} 
\tablehead{
\colhead{Obs. Num} & \colhead{UT Date}  & \colhead{JD$-$2450000}  &
\colhead{Days into Outburst\tablenotemark{a}}}
\startdata
\cutinhead{FAST spectra on 162 dates before the 2000--2002 outburst.}
$\vdots$ & $\vdots$ & $\vdots$ & \\
O1 & 2000 Sep 19 & 1806 & 17 \\
O2 &  2000 Sep 23 & 1810 & 21 \\
O3 &  2000 Oct 04 & 1821 & 32 \\
O4 &  2000 Nov 25 & 1873 & 84 \\
O5\tablenotemark{b} &  2000 Dec 03 & 1882 & 92 \\
O6 &  2000 Dec 29 & 1907 & 117 \\
$\vdots$ & $\vdots$ & $\vdots$ & \\
\cutinhead{Plus FAST spectra on 111 additional dates through the end of 2003.}
\enddata
\tablecomments{See the version of Table~\ref{tab:optews} in the electronic
edition of the Journal for the
full listing of FAST optical spectra.}
\tablenotetext{a}{The beginning of the 2000--2002 outburst was taken
to be JD 2451790.} 
\tablenotetext{b}{From the William Herschel Telescope, courtesy of R. Hynes.}
\end{deluxetable}

\begin{deluxetable}{ccccccccccccccc}
\tablewidth{0pt}
\tabletypesize{\scriptsize}
\tablecaption{Optical Line Equivalent Widths \label{tab:optews}} 
\tablehead{
\colhead{Date} & \colhead{Exposure}  & \colhead{H$\gamma$}  &
\colhead{[O III]} &  \colhead{He I} & \colhead{N III} & \colhead{He II}
& \colhead{H$\beta$} & \colhead{He I} & \colhead{[Fe VII]}
&  \colhead{TiO} & \colhead{H$\alpha$} & \colhead{He I} &
\colhead{Raman} & \colhead{TiO} \\ 
\colhead{(JD)} & \colhead{Time (s)}  & \colhead{4340\AA}  &
\colhead{4363\AA} & \colhead{4388\AA} &
\colhead{4640\AA} & \colhead{4686\AA} & \colhead{4861\AA} & \colhead{5876\AA} & \colhead{6087\AA} &
\colhead{6200\AA} & \colhead{6563\AA} & \colhead{6678\AA} & \colhead{6830\AA} & \colhead{7175\AA}}
\startdata
2449608.87614  &  5.00  & -36.51 &  -8.50 & -3.36 & -7.00 & -79.63 & -44.46
& -7.28  & -1.55  &  0.59 & -120.30 &  -3.96  & -9.69   & 1.02 \\
 2449608.87672 &  5.00 & -37.96 &  -8.25 & -3.21 & -6.42 & -79.45 & -45.29 &
-7.27  & -1.23  &  0.57 & -123.00 &  -3.49 &  -9.64  &  1.01 \\
 2449608.87730 & 60.00  & -38.71  & -9.86  & -4.65 &-7.53 & -81.31 & -45.35
& -7.78  & -1.38  &  0.59 & -79.90  & -3.43 &  -9.56  &  1.01 \\
 2449693.62117 &  5.00  & -52.39  & -3.37  &  -3.11 & -8.99 & -86.84 & -90.62
& -7.80  & -1.88  &  0.55  & -179.72  & -4.18 & -11.34  &  0.86 \\
 2449693.62223 & 30.00 & -50.95 &  -4.87  & -2.97 & -8.63 & -88.37 & -94.11  &
-7.32  & -2.12  &  0.56 & -171.12  & -4.15 & -11.46  &  0.86 \\
 2449725.57784  & 5.00 & -51.76  & -2.92 &  -3.71 & -8.01 & -84.37 & -96.06  &
-7.88  & -1.96 &   0.56 & -190.80  & -5.10 & -10.35 &   0.84 \\
 2449725.57821 & 30.00 & -54.49 &  -3.25 &  -2.84 & -6.67 & -80.26 & -100.16
& -7.82  & -1.79  &  0.58 & -170.60  & -5.03 &  -9.94  &  0.85 \\
 2449725.57922 & 60.00 & -55.38  & -4.08  & -2.60 & -7.41 & -82.00 & -97.73  &
-7.93  & -1.81  &  0.57 & -153.39 &  -5.25 & -10.04  &  0.86 \\
2449749.56896 & 60.00 & -55.03  & -3.93 & -2.90 & -6.53 & -78.05 & -95.47 &
-10.63 &  -1.79  &  0.56 & -54.30 &  -5.32 &  -9.88  &  0.76 \\
2449749.56999 & 60.00 & -57.80  & -3.69 & -2.55 & -7.11 & -73.69 & -95.69 &
-10.13 &  -1.92  &  0.56 & -37.69  & -5.38 & -10.98 &   0.80 \\
$\vdots$ & $\vdots$ & $\vdots$ & $\vdots$ & $\vdots$ & $\vdots$ &
$\vdots$ & $\vdots$ & $\vdots$ & $\vdots$ & $\vdots$ & $\vdots$ &
$\vdots$ & $\vdots$ 
\enddata
\tablecomments{The complete version of this table is in the electronic
edition of the Journal.  The printed edition contains only a sample.
The complete version contains equivalent widths from 
826 spectra between JD 2449608.8761 and JD 2453002.5699.  Equivalent
widths are given in units of \AA\, and have errors of $\sim$ 5\%--10\%.}
\end{deluxetable}

\begin{deluxetable}{clccccc}
\tablewidth{0pt}
\tablecaption{FUV Observations \label{tab:fuv}} 
\tablehead{
\colhead{Obs.} &
\colhead{UT Date}           & \colhead{JD} &
\colhead{Obs}  & \colhead{$f$\tablenotemark{a}} & 
\colhead{$f$\tablenotemark{a}}   &
\colhead{$f$\tablenotemark{a}} \\
\colhead{Num.} & \colhead{} & \colhead{$-$2450000} & \colhead{Length (s)} &
\colhead{(959.5\,\AA)} & \colhead{(1058.7\,\AA)} & \colhead{(1103.4\,\AA)}} 
\startdata
F1 & 2000 Nov 16 & 1865.1 & 8203 & $3.6\pm 0.2$ & $7.3\pm 0.1$ &
$9.5\pm 0.2$\\ 
F2 & 2000 Nov 27 & 1875.8 & 10847 & $8.4\pm 0.6$ & $11.1\pm 0.1$ &
$13.2\pm 0.4$\\ 
F3 & 2000 Dec 15 & 1894.4 & 10444 & $8.2\pm 0.3$ & $11.1\pm 0.3$ &
$13.5\pm 0.03$\\ 
F4 & 2001 Jul 20 & 2111.1 & 10928 & $3.8\pm 0.2$ & $5.4\pm 0.1$ &
$5.8\pm 0.8$\\
F5 & 2001 Jul 22 & 2112.8 & 13081 & $3.8\pm 0.2$ & $5.6\pm 0.1$ &
$5.9\pm 0.2$\\
F6 & 2001 Sep 30 & 2183.1 & 10825 & $3.7\pm 0.3$ & $5.6\pm 0.2$ &
$5.9\pm 0.2$ \\ 
F7 & 2002 Jul 05 & 2460.6 & 3731 & $<$0.2 & $0.70\pm 0.05$ & $0.62\pm 0.05$\\ 
\nodata & 2002 Oct 20 & 2568.5 & 59 & \multicolumn{3}{c}{Safety Snap}
\\ 
F8 & 2002 Oct 22 & 2570.4 & 6736 & $<$0.2 & $0.1\pm 0.07$ & $<$0.2\\
F9 & 2003 Aug 04  & 2856.1 & 14175 & $0.5\pm 0.1$ & $0.78\pm 0.05$ &
$0.72\pm 0.07$ \\
\enddata
\tablenotetext{a}{FUSE flux densities in units of $10^{-13}$ erg
cm$^{-2}$ s$^{-1}$ \AA$^{-1}$.}  
\end{deluxetable}

\begin{deluxetable}{ccccrccc}
\tablewidth{0pt}
\tablecaption{Radio Observations \label{tab:radio}} 
\tablehead{
\colhead{Obs.} &
\colhead{UT Date}           & \colhead{Day\tablenotemark{a}}      &
\colhead{Instrument}          & \colhead{$f$\tablenotemark{c}}  &
\colhead{$f$\tablenotemark{c}} & \colhead{$f$\tablenotemark{c}} &
\colhead{$f$\tablenotemark{c}} \\ 
\colhead{Num} & \colhead{} &\colhead{} &\colhead{(Config\tablenotemark{b})}
& \colhead{(1.4 GHz)} & \colhead{(5 GHz)} & \colhead{(8.5 GHz)} & \colhead{(15 GHz)}} 
\startdata
R1 & 2000 Oct 13 & 1830 & VLA (D) & $<$ 0.29 & 0.42 (0.06) & \nodata & 0.87 (0.13) \\
R2 & 2000 Nov 17 & 1865 & VLA (A) & 0.28 (0.05) & 0.80 (0.03)& \nodata& 2.45 (0.09)\\
R3 & 2001 Jan 12 & 1921 & VLA (A) & $<$ 0.29& 1.21 (0.10) &\nodata & 2.94 (0.50)\\
R4 & 2001 Jan 28 & 1937 & MERLIN & \nodata & 1.32 (0.37)& \nodata & \nodata\\
R5 & 2001 Feb 28 & 1968 & MERLIN & \nodata & 0.70 (0.21) & \nodata &\nodata\\
R6 & 2001 Mar 15 & 1983 & VLA (B) & $<$ 0.17& 1.09 (0.03) & \nodata& 3.50 (0.12)\\
R7 & 2001 May 08  & 2037 & VLA (B) & 0.69 (0.02) & 1.02 (0.02) &
\nodata & 2.53 (0.09)\\
R8 & 2001 Aug 10 & 2131 & VLA (C) & 0.28 (0.08) & 1.06 (0.04)& \nodata&2.14 (0.14)\\
R9 & 2001 Sep 18 & 2170 & MERLIN &
\nodata& 1.20 (0.26)\tablenotemark{d} & \nodata & \nodata \\
R10 & 2001 Oct 23 & 2205 & MERLIN & \nodata & 0.86 (0.20) & \nodata & \nodata\\
R11 & 2002 Apr 30 & 2394 & MERLIN & \nodata & 0.78 (0.20) & \nodata & \nodata \\
R12 & 2002 May 06  & 2400 & MERLIN & \nodata & 0.64 (0.20) & \nodata &\nodata \\
R13 & 2002 Dec 09  & 2618 & VLA (C) & \nodata & 0.50 (0.10) & 0.84 (0.07) & 
\nodata\\
R14 & 2003 Jan 03  & 2643 & VLA(C) & \nodata & 0.42 (0.09) & 0.84 (0.08) & \nodata\\
R15 & 2003 Jul 24  & 2845 & VLA (A)
& \nodata & 0.65 (0.06)\tablenotemark{d} & \nodata & 1.55 (0.16)\\
\enddata
\tablenotetext{a}{Day = JD$-$2450000.}

\tablenotetext{b}{Config = VLA Configuration.  The configuration of the
VLA antennas is changed over time, between the most extended (A)
configuration, with a resolution of about 0.35 arcsec at 4.86 GHz, and
the most compact (D) configuration, with a resolution some 35 times
worse.  See, e.g., \cite{tay04} for details.}

\tablenotetext{c}{Flux densities in mJy.}
\tablenotetext{d}{Spatially resolved.  Flux densities quoted for these 
observations are the total flux densities.}
\end{deluxetable}

\begin{deluxetable}{lcccccc} 
\tablewidth{0pt}
\tablecaption{X-Ray Observations \label{tab:xray}} 
\tablehead{
\colhead{Obs. Num/} & \colhead{UT Date}  & \colhead{JD}  &
\colhead{Orbital} &
\colhead{Obs. Length\tablenotemark{b}}  &  \colhead{Source} \\
\colhead{Inst} & \colhead{} & \colhead{-2450000} &
\colhead{Phase\tablenotemark{a}} & \colhead{(ksec)} &
\colhead{Counts}}
\startdata
X1/$Chandra$ (ACIS-S/HETGS) & 2000 Nov 13 & 1862 & 0.12 & 19 & 
109 \\
X2/$XMM-Newton$ (EPIC-pn) & 2001 Jan 28 & 1938  & 0.21 &
20/16 & 
$\gtrsim$1500 \\
X3/$XMM-Newton$ (EPIC-pn) & 2001 Jun 11 & 2072 & 0.39 &
15/4 & 
$\approx$140 \\
\enddata
\tablenotetext{a}{Orbital phase from ephemeris of \cite{mk96}.}
\tablenotetext{b}{Second quantity for the $XMM-Newton$ observations is the
remaining observation length after removal of high-background data.}
\end{deluxetable}

\begin{deluxetable}{lcccccccc}
\tablewidth{0pt}
\tabletypesize{\footnotesize}
\tablecaption{Best-Fitting Parameters for the X-ray Spectra \label{tab:xfits}}
\tablehead{
\colhead{UT} & \colhead{$N_H$} & \colhead{$\Gamma$} &
\colhead{$kT$} & \colhead{E$_{edge 1}$} & 
\colhead{E$_{edge 2}$} & 
\colhead{$\chi^2_{\nu}/$dof} &
\colhead{$f_{ab}$\tablenotemark{a}} & \colhead{$f_{unab}$\tablenotemark{b}} \\
\colhead{Date}     & \colhead{($10^{21}$ cm$^{-2}$)} & \colhead{} &
\colhead{(keV)} & \colhead{(keV)} &  \colhead{(keV)} & \colhead{} }
\startdata
X1 (2000 Nov)  & $1.4^{+2.5}_{-1.4}$ & \nodata
&$0.21^{+0.04}_{-0.04}$& \nodata &$1.02^{+0.06}_{-0.05}$ &0.53/5 & $2.5\pm0.2$ & $5.1\pm0.5$\\
X2 (2001 Jan) & $1.8^{+1.7}_{-0.9}$ & $0.79^{+2.05}_{-2.01}$& 
$0.11^{+0.02}_{-0.01}$&
$0.64^{+0.02}_{-0.02}$&$0.96^{+0.03}_{-0.03}$&1.11/56& $1.3\pm0.04$ & 
$3.1\pm0.1$\\
X3 (2001 June) & $2.4^{+4.8}_{-2.4}$ & \nodata &
$0.12^{+0.07}_{-0.07}$& \nodata & \nodata & 0.35/11 
& $0.2\pm0.05$ & $1.4\pm0.3$\\
\enddata
\tablecomments{All fit parameters are for blackbody models plus
absorption edges, with an additional powerlaw component in observation
X2 ($\Gamma$ is the photon index).  The quoted uncertainties are 90\% confidence.}
\tablenotetext{a}{0.3--7 keV absorbed flux ($10^{-13}$ erg cm$^{-2}$ s$^{-1}$).}
\tablenotetext{b}{0.3--7 keV unabsorbed flux ($10^{-13}$ erg cm$^{-2}$ s$^{-1}$).}
\end{deluxetable}

\begin{deluxetable}{clcccccc} 
\tablewidth{0pt}
\tablecaption{Hot-Component Luminosities from FUV
Fluxes\label{tab:fuvlums}} 
\tablehead{
\colhead{Obs.} & \colhead{UT Date} & \colhead{JD} &
\colhead{$T_{hot}$ }  &  \colhead{$F_{\lambda 1059}$\tablenotemark{a}} &
\colhead{$F_{\lambda 1103}$\tablenotemark{a}} & \colhead{$R_{hot}$ } &
\colhead{$L_{hot}$ } \\
\colhead{Num.} & \colhead{} & \colhead{$-$2450000} & \colhead{($10^3$
K)} & \colhead{} & \colhead{} &
\colhead{($\rsun$)} & \colhead{($\lsun$)}}  
\startdata
F1 & 2000 Nov 16 & 1865.1 & 92$\pm 20$ & 250 & 240 & $0.32\pm 0.05$ &
6600$\pm 1700$ \\ 
F2 & 2000 Nov 27 & 1875.8 & 94$\pm 20$ & 380 & 330 & $0.36\pm 0.06$ &
9400$\pm 2500$\\
F3 & 2000 Dec 15 & 1894.4 & 95$\pm 20$ & 380 & 340 & $0.36\pm 0.06$ &
9800$\pm 2500$ \\
F4 & 2001 Jul 20 & 2111.1 & 
120$\pm 20$ & 180 & 150 & $0.18\pm 0.03$ & 6100$\pm 1500$\\
F5 & 2001 Jul 22 & 2112.8 & 
120$\pm 20$ & 190 & 150 & $0.18\pm 0.03$ & 6100$\pm 1500$\\
F6 & 2001 Sep 30 & 2183.1 & 
120$\pm 20$ & 190 & 150 & $0.17\pm 0.03$ & 5600$\pm 1400$\\ 
F7\tablenotemark{b} & 2002 Jul 5 & 2460.6 & 150$\pm 30$ & 24 & 16 & $0.06\pm 0.01$ &
1500$\pm 400$\\
%
%
F8\tablenotemark{b} & 2002 Oct 22 & 2570.4 & 
120$\pm 25$ & 3 & $<$5 & $0.03\pm 0.01$ & 200$\pm 50$\\
F9\tablenotemark{b} & 2003 Aug 04  & 2856.1 & 160$\pm 35$ & 26 & 18 &
$0.06\pm 0.01$ & 2000$\pm 500$ \\
\enddata
\tablenotetext{a}{$FUSE$ extinction-corrected flux densities at
1058.7\, \AA\, and 1103.4\, \AA\, in units
of $10^{-13}$ erg cm$^{-2}$ s$^{-1}$ A$^{-1}$.  Fluxes are corrected for
interstellar extinction using the extinction curve of \cite{ccm89} as
implemented in the IDL routine `unred\_ccm' (see
http://archive.stsci.edu/pub/iue/software/iuedac/procedures/unred\_ccm.pro), with $E(B-V) = 0.27$ mag and $R=
A(V)/E(B-V) = 3.1$. 
}
\tablenotetext{b}{For observations F7--F9, the flux ratios $F_{\lambda
1059}/F_{\lambda 1103}$ are unphysical for a WD photosphere.  They
have therefore likely been contaminated by non-photospheric absorption
and/or emission.  The formal values of $R_{hot}$ and $L_{hot}$ for
these observations should thus be used with extreme caution.}
\end{deluxetable}

\end{document}